\makeatletter
\@ifundefined{@parse@version@dash}{%
\def\@parse@version#1{\@parse@version@0#1}
\def\@parse@version@#1/#2/#3#4#5\@nil{%
\@parse@version@dash#1-#2-#3#4\@nil}
\def\@parse@version@dash#1-#2-#3#4#5\@nil{%
  \if\relax#2\relax\else#1\fi#2#3#4 }
}{}
\makeatother

\documentclass[twocolumn, amssymb, nobibnotes, aps, pra, superscriptaddress, floatfix, longbibliography]{revtex4-2}
\usepackage[pdftex]{graphicx}
\usepackage{amsmath, amssymb}
\usepackage{braket}
\usepackage{natbib}
\usepackage{color}
\usepackage[caption=false]{subfig}

\newcommand{\bs}[1]{\boldsymbol{#1}}
\newcommand{\nm}{\nonumber}
\newcommand{\pdif}[2]{\dfrac{\partial #1}{\partial #2}}

\begin{document}

% Use the \preprint command to place your local institutional report
% number in the upper righthand corner of the title page in preprint mode.
% Multiple \preprint commands are allowed.
% Use the 'preprintnumbers' class option to override journal defaults
% to display numbers if necessary
%\preprint{}

%Title of paper
\title{Variational quantum algorithm based on the minimum potential energy for solving the Poisson equation}

% repeat the \author .. \affiliation  etc. as needed
% \email, \thanks, \homepage, \altaffiliation all apply to the current
% author. Explanatory text should go in the []'s, actual e-mail
% address or url should go in the {}'s for \email and \homepage.
% Please use the appropriate macro foreach each type of information

% \affiliation command applies to all authors since the last
% \affiliation command. The \affiliation command should follow the
% other information
% \affiliation can be followed by \email, \homepage, \thanks as well.
\author{Yuki Sato}
\email[]{yuki-sato@mosk.tytlabs.co.jp}
%\homepage[]{Your web page}
%\thanks{}
%\altaffiliation{}
\affiliation{Toyota Central R\&D Labs., Inc., Koraku Mori Building 10F, 1-4-14 Koraku, Bunkyo-ku, Tokyo 112-0004, Japan}

\author{Ruho Kondo}
%\email[]{Your e-mail address}
%\homepage[]{Your web page}
%\thanks{}
\affiliation{Toyota Central R\&D Labs., Inc., 41-1, Yokomichi, Nagakute, Aichi 480-1192, Japan}

\author{Satoshi Koide}
%\email[]{Your e-mail address}
%\homepage[]{Your web page}
%\thanks{}
%\altaffiliation{}
\affiliation{Toyota Central R\&D Labs., Inc., Koraku Mori Building 10F, 1-4-14 Koraku, Bunkyo-ku, Tokyo 112-0004, Japan}

\author{Hideki Takamatsu}
%\email[]{Your e-mail address}
%\homepage[]{Your web page}
%\thanks{}
%\altaffiliation{}
\affiliation{Toyota Motor Corporation, 1 Toyota-Cho, Toyota, Aichi 471-8571, Japan}

\author{Nobuyuki Imoto}
%\email[]{Your e-mail address}
%\homepage[]{Your web page}
%\thanks{}
%\altaffiliation{}
\affiliation{Institute for Photon Science and Technology, School of Science, The University of Tokyo, 7-3-1, Hongo, Bunkyo-ku, Tokyo 113-0033, Japan}

%Collaboration name if desired (requires use of superscriptaddress
%option in \documentclass). \noaffiliation is required (may also be
%used with the \author command).
%\collaboration can be followed by \email, \homepage, \thanks as well.
%\collaboration{}
%\noaffiliation

\date{\today}

\begin{abstract}
% insert abstract here
Computer-aided engineering techniques are indispensable in modern engineering developments. 
In particular, partial differential equations are commonly used to simulate the dynamics of physical phenomena, but very large systems are often intractable within a reasonable computation time, even when using supercomputers. 
To overcome the inherent limit of classical computing, we present a variational quantum algorithm for solving the Poisson equation that can be implemented in noisy intermediate-scale quantum devices. 
The proposed method defines the total potential energy of the Poisson equation as an expectation of certain observables, which are decomposed into a linear combination of Pauli operators and simple observables.
The expectation value of these observables is then minimized with respect to a parameterized quantum state. 
Because the number of decomposed terms is independent of the size of the problem, this method requires relatively few quantum measurements. 
Numerical experiments demonstrate the faster computing speed of this method compared with classical computing methods and a previous variational quantum approach. 
We believe that our approach brings quantum computer-aided techniques closer to future applications in engineering developments.
Code is available at https://github.com/ToyotaCRDL/VQAPoisson.
\end{abstract}

% insert suggested keywords - APS authors don't need to do this
%\keywords{}

%\maketitle must follow title, authors, abstract, and keywords
\maketitle

% body of paper here - Use proper section commands
% References should be done using the \cite, \ref, and \label commands
%\section{}
% Put \label in argument of \section for cross-referencing
%\section{\label{}}

%\subsection{}
%\subsubsection{}

\section{Introduction}
Partial differential equations (PDEs) are frequently used to describe the dynamics of physical phenomena such as heat conduction, fluid dynamics, and solid mechanics \cite{Renardy2004introduction}.
Solving these problems as quickly as possible is a key factor in accelerating engineering developments that combine computer simulations and real-world experiments.
Although many advances have been made in terms of treating very large physical systems using classical computers \cite{Klawonn2015, Toivanen2018}, obtaining solutions within a reasonable computational time is increasingly intractable.
State-of-the-art computational techniques can perform simulations of systems with up to tens of billions of degrees of freedom using Fugaku, which is one of the most powerful supercomputers; these simulations typically take several hours \cite{Kato2020toward, Fujita2021high}.

Another possible and attractive approach that would significantly reduce the computational costs of solving PDEs is the application of quantum computing.
Quantum computing has attracted considerable attention over recent decades as a potential means of providing faster and more powerful computation than classical computing.

Quantum algorithms for solving linear systems have been developed \cite{Harrow2009quantum, Childs2017quantum}, and these provide an exponential speedup over classical algorithms when the coefficient matrix of the linear system is sparse.
Cao \textit{et al.} \cite{Cao2013quantum} developed an algorithm for solving the Poisson equation, which is one of the most important PDEs in various areas of engineering, such as electrostatics \cite{Griffiths1999introduction} and computational fluid dynamics \cite{Chung2010computational, Blazek2015computational}.
Although it is widely believed that the abovementioned quantum algorithms will demonstrate quantum supremacy once fault-tolerant quantum computers with sufficient qubits and error correction techniques are ready, it is expected to be a long time until such a quantum computer can be realized.
Thus, recent interest in quantum computing has focused on the development of quantum algorithms that perform some meaningful computations with a small number of qubits and a shallow circuit.
In other words, it is important to construct a quantum algorithm that can be implemented on so-called noisy intermediate-scale quantum (NISQ) devices \cite{Preskill2018quantum}.

Variational quantum algorithms (VQAs) \cite{Cerezo2021variational} are possible candidates for use on NISQ devices, as they exhibit several advantages over classical algorithms.
In VQAs, a certain cost function is written as a function of the expectation value of some observables, and this function is evaluated on a quantum computer using a trial quantum state prepared by a parameterized quantum circuit.
This cost function evaluation is iteratively performed while the classical parameters are updated so as to minimize the cost function.
Thus, VQAs are effectively hybrid quantum--classical algorithms.
A well-known example of a VQA is a variational quantum eigensolver (VQE) \cite{Peruzzo2014variational} in which the cost function is the expectation of the system Hamiltonian.
The VQE was originally developed to calculate the lowest eigenvalue of a system Hamiltonian via parameterized quantum circuits, also called ansatz \cite{Fedorov2021vqe}, and has been widely studied in terms of the quantum circuits suitable for hardware architectures \cite{Kandala2017hardware, Ganzhorn2019gate} and the construction of efficient optimizers \cite{Jones2019variational, Higgott2019variational, Yuan2021hybrid}.
Another popular VQA is the quantum approximate optimization algorithm (QAOA) \cite{Farhi2014quantum}, which can be used to solve combinational optimization problems.
The QAOA was originally based on the concept of adiabatic quantum computing \cite{Farhi2000quantum, Childs2001robustness}; it has since been extended to more general families of operators, and is thus widely known as the quantum alternating operator ansatz \cite{Hadfield2019quantum}.

VQAs for solving linear systems have also been proposed.
Bravo-Prieto \textit{et al.} \cite{Bravo2020variational} proposed a VQA for a linear system in which the system matrix is written as a linear combination of unitary operators.
This algorithm prepares a quantum state whose amplitude is proportional to the solution of a linear system output by the Harrow--Hassidim--Lloyd algorithm \cite{Harrow2009quantum}.
Liu \textit{et al.} \cite{Liu2020variational} presented a scheme for efficiently solving the Poisson equation by explicitly decomposing a system matrix derived from the finite difference discretization of the Poisson equation. 
This scheme requires $\mathcal{O}(\text{log}N)$ expectation calculations at every iteration in the optimization procedure, where $N$ is the size of the problem.
However, these methods provide normalized solutions and do not explicitly provide the norm of the solutions.
In the field of classical physics, the unknown quantities of PDEs directly describe physical quantities whose norms are not necessarily equal to $1$.
For example, the temperature field in steady-state heat conduction is governed by the Poisson equation, and we need to know the scale of the temperature when assessing whether the temperature in a certain device exceeds the operating temperature limit.
In an engineering sense, therefore, the norms of these physical quantities provide important information.

This paper describes a method for solving the Poisson equation by formulating the cost function based on the concept of the minimum potential energy.
This approach naturally leads to an expression for the norm of the solutions.
The system matrix of the discretized Poisson equation is split into a linear combination of the tensor product of Pauli operators and simple observables.
The number of decomposed terms is independent of the system size, i.e., $\mathcal{O}(1)$, which significantly reduces the required number of expectation calculations on quantum computers.

The remainder of this paper is organized as follows.
In Sec.~\ref{sec:formula}, the optimization problem for solving the Poisson equation is formulated based on the concept of VQAs.
Section~\ref{sec:methods} describes the numerical implementation based on this formulation, and then Sec.~\ref{sec:numel} provides several results from numerical experiments. 
Finally, the conclusions to this study are presented in Sec.~\ref{sec:conc}.

\section{Formulation} \label{sec:formula}
\subsection{Derivation of optimization problem}
Consider the Poisson equation defined in an open bounded domain $\Omega \subset \mathbb{R}^d$, where $d$ is the number of spatial dimensions.
Let $u(\bs{x})$ denote the state field at the spatial coordinate $\bs{x} \in \mathbb{R}^d$.
Consider a function $f(\bs{x}): \Omega \rightarrow \mathbb{C}$. The Poisson equation is then defined as
\begin{equation}
	- \nabla^2 u(\bs{x}) = f(\bs{x}) \quad \text{in} \quad \Omega.
\end{equation}
where $\nabla^2$ is the Laplace operator.
As typical boundary conditions, the Neumann and Dirichlet boundary conditions are, respectively, defined as
\begin{align}
	& - \bs{n} \cdot \nabla u(\bs{x}) = 0 \quad \mathrm{on} \quad \Gamma_\mathrm{N}, \\
	& u(\bs{x}) = 0 \quad \text{on} \quad \Gamma_\mathrm{D},
\end{align}
where $\Gamma_\mathrm{N}$ and $\Gamma_\mathrm{D}$ denote the boundaries on which the Neumann and Dirichlet boundary conditions are, respectively, imposed, 
$\bs{n}$ is the normal vector pointing outward from the boundary of the domain $\Omega$, and $\Gamma_\mathrm{N} \cup \Gamma_\mathrm{D} = \partial \Omega$, which means that any point on $\partial \Omega$ is included in either $\Gamma_\mathrm{N}$ or $\Gamma_\mathrm{D}$.
In the following, the spatial coordinate $\bs{x}$ is omitted for readability.

Let us consider the total potential energy, defined as
\begin{equation}
	E := \dfrac{1}{2} \int_{\Omega} \nabla v^\ast \cdot \nabla v ~d\Omega - \dfrac{1}{2} \int_{\Omega} v^\ast f ~d\Omega - \dfrac{1}{2} \int_{\Omega} f^\ast v ~d\Omega, \label{eq:E}
\end{equation}
for $v \in \mathcal{V}$ with
\begin{equation}
	\mathcal{V} := \{ v \in H^1(\Omega) ~|~ v = 0 ~ \mathrm{on} ~ \Gamma_\mathrm{D} \}
\end{equation}
where $H^1(\Omega)$ is a Sobolev space.
The stationary condition of the total potential energy with respect to a function $v$ yields
\begin{align}
	0 =& dE(v; \delta v) \nm \\
	=& \dfrac{1}{2} \int_{\Omega} \nabla \delta v^\ast \cdot \nabla v ~d\Omega + \dfrac{1}{2} \int_{\Omega} \nabla v^\ast \cdot \nabla \delta v ~d\Omega \nm \\
	&- \dfrac{1}{2} \int_{\Omega} \delta v^\ast f ~d\Omega - \dfrac{1}{2} \int_{\Omega} f^\ast \delta v ~d\Omega \nm \\
    =& \dfrac{1}{2} \int_{\Gamma_\mathrm{N}} \delta v^\ast \bs{n} \cdot \nabla v ~d\Gamma - \dfrac{1}{2} \int_{\Omega} \delta v^\ast \left( \nabla^2 v + f \right) ~d\Omega \nm \\
    &+ \dfrac{1}{2} \int_{\Gamma_\mathrm{N}} \delta v \left( \bs{n} \cdot \nabla v\right)^\ast ~d\Gamma - \dfrac{1}{2} \int_{\Omega} \delta v \left( \nabla^2 v + f \right)^\ast ~d\Omega \label{eq:dE}
\end{align}
where $dE(v; \delta v)$ represents the G\^{a}teaux derivative of $E$ with respect to $v$ in the direction $ \delta v \in \mathcal{V}$.
In Eq.~(\ref{eq:dE}), Gauss's theorem was applied from the second line to the third, and the integration on $\Gamma_\mathrm{D}$ vanished because the trace of $\delta v = 0$ on the boundary where the Dirichlet boundary condition is imposed.
Because $\delta v$ is arbitrary, the above stationary condition recovers the Poisson equation from the second and fourth terms of the right-hand side of Eq.~(\ref{eq:dE}), which is known as the principle of minimum potential energy.
This means that minimizing the total potential energy with respect to the function $v$ yields the state field $u$, governed by the Poisson equation.

Now, to derive the cost function, Eq.~(\ref{eq:E}) is discretized using some technique such as the finite difference or finite element method.
The discretized version of the total potential energy, $E_h$, can be written as
\begin{equation}
	E_h := \dfrac{1}{2} \bs{v}^\ast \cdot A \bs{v} - \dfrac{1}{2} \bs{v}^\ast \cdot \bs{f} - \dfrac{1}{2} \bs{f}^\ast \cdot \bs{v} \label{eq:Eh},
\end{equation}
where $\bs{v}$ and $\bs{f}$ denote vectors with component values of $v$ and $f$, respectively, at the nodes discretizing the domain $\Omega$.
$A$ is a positive-definite symmetric matrix obtained from the discretization of the first term in Eq.~(\ref{eq:E}) into the quadratic form.
Let $\Ket{v}$ and $\Ket{f}$ be the state vectors encoding $\bs{v}$ and $\bs{f}$, respectively.
Without loss of generality, the squared norm of $\bs{f}$ can be assumed to be $1$ from the linearity of the Poisson equation, which means that $\Ket{f}$ can be regarded as a quantum state.
However, it should be noted that $\Ket{v}$ is not necessarily a quantum state, because the squared norm of the solution vector is not necessarily $1$.
To represent the solution vector using a parameterized quantum state $\Ket{\psi(\bs{\theta})}$ with parameter vectors $\bs{\theta}$, we thus introduce a parameter $r \in \mathbb{R}$ and then represent $\Ket{v} = r \Ket{\psi(\bs{\theta})}$.
Substituting $\Ket{v} = r \Ket{\psi(\bs{\theta})}$ and $\Ket{f}$, respectively, for $\bs{v}$ and $\bs{f}$ in Eq.~(\ref{eq:Eh}), we obtain
\begin{align}
	E_h =& \dfrac{1}{2} r^2 \Braket{\psi(\bs{\theta}) | A | \psi(\bs{\theta})} \nm \\
	&- \dfrac{1}{2} r \Braket{\psi(\bs{\theta}) | f} - \dfrac{1}{2} r \Braket{f | \psi(\bs{\theta})}.
\end{align}
Introducing the quantum state $\Ket{f, \psi(\bs{\theta})} := \left( \Ket{0}\Ket{f} + \Ket{1}\Ket{\psi(\bs{\theta})} \right) / \sqrt{2}$, the discretized version of the total potential energy $E_h$ in Eq.~(\ref{eq:Eh}) can be described as
\begin{align}
	E_h(r, \bs{\theta}) =& \dfrac{1}{2} r^2 \Braket{\psi(\bs{\theta}) | A | \psi(\bs{\theta})} \nm \\
	&- r \Braket{f, \psi(\bs{\theta}) | X \otimes I^{\otimes n} | f, \psi(\bs{\theta})}, \label{eq:Eh_quantum}
\end{align}
where $I = \Ket{0}\Bra{0} + \Ket{1}\Bra{1}$, $X = \Ket{0}\Bra{1} + \Ket{1}\Bra{0}$, and $n = \log_2 N$, where $N$ is the number of nodes.
Because $r$ and $\bs{\theta}$ parameterize $\Ket{v}$, minimizing the total potential energy with respect to $\Ket{v}$ corresponds to the minimization of $E_h(r, \bs{\theta})$ in Eq.~(\ref{eq:Eh_quantum}) with respect to $r$ and $\bs{\theta}$.
The optimization problem that solves the Poisson equation can, therefore, be formulated as
\begin{equation}
    \min_{r, \bs{\theta}} \quad E_h(r, \bs{\theta}). \label{eq:opt}
\end{equation}
Letting $r_\mathrm{opt}$ and $\bs{\theta}_\mathrm{opt}$ denote the solution of this optimization problem, the state vector that encodes the solution of the Poisson equation is $\Ket{u} = r_\mathrm{opt} \Ket{\psi(\bs{\theta}_\mathrm{opt})}$.
Here, these two minimizations with respect to $r$ and $\bs{\theta}$ can be performed sequentially, i.e., minimizing $E_h$ with respect to $r$, followed by $\bs{\theta}$ as
\begin{equation}
    \min_{r, \bs{\theta}} \quad E_h(r, \bs{\theta}) = \min_{\bs{\theta}} \quad E_h(r_\mathrm{opt}(\bs{\theta}), \bs{\theta}),
\end{equation}
where $r_\mathrm{opt}(\bs{\theta})$ is the optimal value of $r$ for a fixed value of $\bs{\theta}$.
As the cost function $E_h$ is parabolic with respect to $r$, the optimal solution of $r$ for a given $\theta$ is analytically derived by the following necessary condition for optimality, which requires that the partial derivative of $E_h(r, \bs{\theta})$ with respect to $r$ is equal to $0$:
\begin{align}
    0 &= \pdif{E_h(r, \bs{\theta})}{r} \nm \\
	 &= r \Braket{\psi(\bs{\theta}) | A | \psi(\bs{\theta})} - \Braket{f, \psi(\bs{\theta}) | X \otimes I^{\otimes n} | f, \psi(\bs{\theta})}.
\end{align}
Then, we have
\begin{equation}
    r_\mathrm{opt}(\bs{\theta}) = \dfrac{ \Braket{f, \psi(\bs{\theta}) | X \otimes I^{\otimes n} | f, \psi(\bs{\theta})} }{ \Braket{\psi(\bs{\theta}) | A  | \psi(\bs{\theta})} }. \label{eq:norm}
\end{equation}
Consequently, $r$ can be deleted from the cost function, and the optimization problem is to minimize the following Eq.~(\ref{eq:E_h_theta}) with respect to $\bs{\theta}$.
\begin{align}
    E_h(r_\mathrm{opt}(\bs{\theta}), \bs{\theta}) &=-\dfrac{1}{2} \dfrac{ \left( \Braket{f, \psi(\bs{\theta}) | X \otimes I^{\otimes n} | f, \psi(\bs{\theta})} \right)^2 }{ \Braket{\psi(\bs{\theta}) | A  | \psi(\bs{\theta})} }. \label{eq:E_h_theta}
\end{align}
Note that the denominator $\Braket{\psi(\bs{\theta}) | A  | \psi(\bs{\theta})}$ is positive for arbitrary quantum states $\Ket{\psi(\bs{\theta})}$ owing to the positive-definiteness of the operator $A$.

\subsection{Evaluation of cost function} \label{sec:cost_eval}
In this paper, we focus on the one-dimensional Poisson equation, discretized by the finite element method (FEM) \cite{Hughes2012finite}, in which the mesh size of all finite elements is $1$.
For $N$ nodes in one dimension, the matrix $A$ is described using first-order elements, depending on the boundary conditions, as follows:
With periodic boundary conditions, 
\begin{equation}
	A_\text{periodic} := \begin{bmatrix}
		2  & -1 & 0 &  & \ldots &  & 0 & -1 \\
		-1 & 2 & -1 & 0 & \ldots &  &  & 0 \\
		0 & -1 & 2 & -1 & \ldots &  &  & 0 \\
		\vdots &  &  & \ddots &  &  &  & \vdots \\
		0 &  &  & \ldots & 0 & -1 & 2 & -1 \\
		-1 & 0 &  & \ldots &  & 0 & -1 & 2 \\
	\end{bmatrix} \in \mathbb{R}^{N \times N}. \label{eq:A_periodic}
\end{equation}
With Dirichlet boundary conditions, 
\begin{equation}
	A_\text{Dirichlet} := \begin{bmatrix}
		2  & -1 & 0 &  & \ldots &  &  & 0 \\
		-1 & 2 & -1 & 0 & \ldots &  &  & 0 \\
		0 & -1 & 2 & -1 & \ldots &  &  & 0 \\
		\vdots &  &  & \ddots &  &  &  & \vdots \\
		0 &  &  & \ldots & 0 & -1 & 2 & -1 \\
		0 &  &  & \ldots &  & 0 & -1 & 2 \\
	\end{bmatrix} \in \mathbb{R}^{N \times N}. \label{eq:A_D}
\end{equation}
With Neumann boundary conditions, 
\begin{equation}
	A_\text{Neumann} := \begin{bmatrix}
		1  & -1 & 0 &  & \ldots &  &  & 0 \\
		-1 & 2 & -1 & 0 & \ldots &  &  & 0 \\
		0 & -1 & 2 & -1 & \ldots &  &  & 0 \\
		\vdots &  &  & \ddots &  &  &  & \vdots \\
		0 &  &  & \ldots & 0 & -1 & 2 & -1 \\
		0 &  &  & \ldots &  & 0 & -1 & 1 \\
	\end{bmatrix} \in \mathbb{R}^{N \times N}. \label{eq:A_N}
\end{equation}
The matrix $A_\text{Dirichlet}$ is the same as that derived in previous research~\cite{Liu2020variational} using the finite difference method (FDM).
In one dimension, the matrices derived from the FEM using first-order elements coincide with those from the central difference scheme in the FDM.

While previous research~\cite{Liu2020variational} has shown that the matrix $A_\text{Dirichlet}$ can be decomposed into $\mathcal{O}(n)$ terms consisting of Pauli and simple operators, the present study provides the decomposition of the above matrices $A_\text{periodic}$, $A_\text{Dirichlet}$, and $A_\text{Neumann}$ into $\mathcal{O}(1)$ terms consisting of Pauli operators and simple observables.
The matrix $A_\text{periodic}$ can be split into two matrices as follows:
\begin{equation}
A_\text{periodic} = A_{\mathcal{T}_\text{even}} + A_{\mathcal{T}_\text{odd}}, \label{eq:A_decomposition}
\end{equation}
where
\begin{align}
A_{\mathcal{T}_\text{even}} &:= 
\begin{bmatrix}
1  & -1 & 0 & 0 & & \ldots &  & 0 \\
-1 & 1 & 0 & 0 & & \ldots &  & 0 \\
0 & 0 & 1 & -1 & & \ldots &  & 0 \\
0 & 0 & -1 & 1 & & \ldots & & 0 \\
\vdots &  &  & \ddots &  &  &  & \vdots \\
0 &  &  & \ldots & 0 & 0 & 1 & -1 \\
0 &  &  & \ldots & 0 & 0 & -1 & 1 \\
\end{bmatrix}, \label{eq:A_even} \\
A_{\mathcal{T}_\text{odd}} &:= 
\begin{bmatrix}
1  & 0 & 0 &  & \ldots &  & 0 & -1 \\
0 & 1 & -1 & 0 & \ldots &  &  & 0 \\
0 & -1 & 1 & 0 & \ldots &  &  & 0 \\
\vdots &  &  & \ddots &  &  &  & \vdots \\
0 &  &  &  & 0 & 1 & -1 & 0 \\
0 &  &  & \ldots & 0 & -1 & 1 & 0 \\
-1 & 0 &  & \ldots &  & 0 & 0 & 1 \\
\end{bmatrix}. \label{eq:A_odd}
\end{align}

Because $A_{\mathcal{T}_\text{even}}$ can be written in simple form using Pauli operators as 
\begin{equation}
A_{\mathcal{T}_\text{even}} = I^{\otimes {n-1}} \otimes (I-X),
\end{equation}
and $A_{\mathcal{T}_\text{even}}$ and $A_{\mathcal{T}_\text{odd}}$ are interchanged by shifting the node number towards $+1$, $A_{\mathcal{T}_\text{odd}}$ can also be written using Pauli operators as 
\begin{equation}
A_{\mathcal{T}_\text{odd}} = P^{-1} \left( I^{\otimes {n-1}} \otimes (I-X) \right) P,
\end{equation}
where $P$ is a shift operator defined as
\begin{equation}
P := \sum_{i \in [0, 2^n-1]} \Ket{(i+1)~\mathrm{mod}~2^n} \Bra{i}.
\end{equation}
Considering the differences between matrices depending on the three types of boundary conditions in Eqs.~(\ref{eq:A_periodic})--(\ref{eq:A_N}),
these matrices can now be described as follows using the tensor products of the Pauli operators, a shift operator, and a simple Hermitian $I_0= \Ket{0} \Bra{0}$:
\begin{align}
	A_\text{periodic} &= I^{\otimes {n-1}} \otimes (I-X) \nm \\
	                  & \quad + P^{-1} \left( I^{\otimes {n-1}} \otimes (I-X) \right) P \label{eq:A_pauli_periodic} \\
	A_\text{Dirichlet} &= A_\text{periodic} + P^{-1} \left( I_0^{\otimes {n-1}} \otimes X \right) P \label{eq:A_pauli_D}\\
	A_\text{Neumann} &= A_\text{periodic} - P^{-1} \left( I_0^{\otimes {n-1}} \otimes (I-X) \right) P \label{eq:A_pauli_N}.
\end{align}
The expectation values of observables including a shift operator can be evaluated as follows by applying the shift operator, which is unitary, to the quantum state:
\begin{equation}
	\Braket{\phi | P^{-1} H P | \phi} = \Braket{\phi' | H | \phi'},
\end{equation}
where $\Ket{\phi}$ is an arbitrary $n$-qubit state and $\Ket{\phi'} = P \Ket{\phi}$.
Now, the denominator of Eq.~(\ref{eq:E_h_theta}) is described using a linear combination of the Pauli operators and simple Hermitians $I_0$, $I_1$, and the numerator of Eq.~(\ref{eq:E_h_theta}) is originally expressed by the Pauli operators.
Because the expectation values of the identity operator are $1$, i.e.,
\begin{align}
    \Braket{\phi | I^{\otimes n} |\phi} &= 1 \\
    \Braket{\phi | P^{-1} I^{\otimes n} P |\phi} &= \Braket{\phi' | I^{\otimes n} |\phi'} = 1,
\end{align}
the number of terms to be evaluated by a quantum computer is three for the periodic boundary condition, four for the Dirichlet boundary condition, and five for the Neumann boundary condition; of these, one term is for the numerator of Eq.~(\ref{eq:E_h_theta}) and the others are for the linear combination in the denominator of Eq.~(\ref{eq:E_h_theta}).

The above discussion can easily be extended to $d$-dimensional problems using the FDM and FEM.
When using the FDM, the cost function is defined by the $d-$dimensional system matrix $A_d$, that is,
\begin{equation}
A_d := A \otimes I^{\otimes (d-1)} + I \otimes A \otimes I^{\otimes (d-2)} + \cdots + I^{\otimes (d-1)} \otimes A,
\end{equation}
instead of $A$ \cite{Cao2013quantum, Liu2020variational}.
This increases the number of terms to be measured by a factor of $d$ compared with the one-dimensional case, i.e., the number of terms to be measured is $\mathcal{O}(d)$.
When using the FEM, the decomposition of $A_d$ into a linear combination of Pauli operators is derived by defining a graph corresponding to the finite elements, which is explained in Appendix~\ref{sec:graph}.

\section{Implementation} \label{sec:methods}
\subsection{Overview of the algorithm} \label{sec:algo}
Here, we briefly describe the proposed algorithm.

\begin{enumerate}
    \item [Step 1] Initialize a set of parameters $\bs{\theta}$ in a classical computer.
    \item [Step 2] Evaluate the cost function $E_h$ in Eq.~(\ref{eq:E_h_theta}) using a quantum computer.
    \item [Step 3] If a certain terminal condition is satisfied, the optimization procedure halts; otherwise, proceed to Step 4.
    \item [Step 4] Update the set of parameters using some classical optimization scheme, then return to Step 2.
\end{enumerate}

We use several kinds of terminal conditions in the numerical experiments. 
These will be specified in Sec. \ref{sec:numel}.

\subsection{State preparation} \label{sec:state_preparation}
\begin{figure}
\includegraphics[width=5cm]{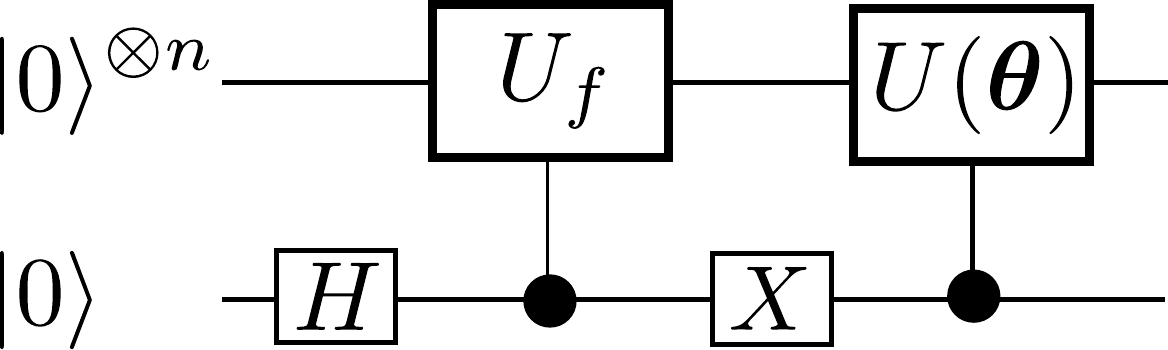}
\caption{Quantum circuit for preparing $\Ket{f, \psi(\bs{\theta})} := \left( \Ket{0}\Ket{f} + \Ket{1}\Ket{\psi(\bs{\theta})} \right) / \sqrt{2}$. The leftmost qubit is shown in the bottom line of the circuit. \label{fig:c_ansatz}}
\end{figure}

In the proposed method, the quantum state $\Ket{\psi (\bs{\theta})}$ is prepared by applying a sequence of parameterized quantum gates $U(\bs{\theta})$, the so-called ansatz, to the $\Ket{0}^{\otimes n}$ state.
We use a hardware-efficient ansatz \cite{Kandala2017hardware}, specifically an alternating layered ansatz consisting of $R_Y$ gates and controlled $Z$ gates \cite{Bravo2020variational}, to constrain the amplitude in the real space.
This is valid in the solution of the Poisson equation with $f(\bs{x}) \in \mathbb{R}$.
In the state-preparation stage, the state vector $\Ket{f}$, which corresponds to the source term of the Poisson equation, must also be prepared.
To encode arbitrary state vectors, amplitude encoding techniques are required \cite{Mottonen2004transformation, Iten2016quantum, Araujo2021divide}.
These encode classical data into the amplitudes of a quantum state.
In the current study, for simplicity, we assume that there is a unitary $U_f$ that efficiently prepares $\Ket{f}$ from $\Ket{0}^{\otimes n}$ (i.e., $\Ket{f} = U_f \Ket{0}^{\otimes n}$).
The quantum state $\Ket{f, \psi (\bs{\theta})}$ for evaluating the numerator of the cost function in Eq.~(\ref{eq:E_h_theta}) is prepared using an auxiliary qubit and controlled versions of the parameterized quantum circuit $U(\bs{\theta})$ and the unitary $U_f$ of the quantum circuit shown in Fig.~\ref{fig:c_ansatz}, which is more expensive than circuits preparing $\Ket{\psi (\bs{\theta})}$ and $\Ket{f}$.
Note that when we constrain the amplitudes of $\Ket{\psi (\bs{\theta})}$ and $\Ket{f}$ in the real space, the numerator of the cost function can also be evaluated without the quantum state $\Ket{f, \psi (\bs{\theta})}$ as follows:
\begin{align}
& \left( \Braket{f, \psi(\bs{\theta}) | X \otimes I^{\otimes n} | f, \psi(\bs{\theta})} \right)^2 \nm \\
&= \Braket{\psi (\bs{\theta}) | f} \Braket{f | \psi (\bs{\theta})} \quad (\text{if } \mathrm{Im}(\Braket{\psi (\bs{\theta}) | f})=0) \nm \\
&= \braket{\psi (\bs{\theta}) | U_f | 0} \braket{0 |U_f^\dagger | \psi (\bs{\theta})},
\end{align}
where the first equality holds when the amplitudes of $\Ket{\psi (\bs{\theta})}$ and $\Ket{f}$ are in the real space.

\subsection{Quantum circuit for the shift operator}
\begin{figure}
\includegraphics[width=5cm]{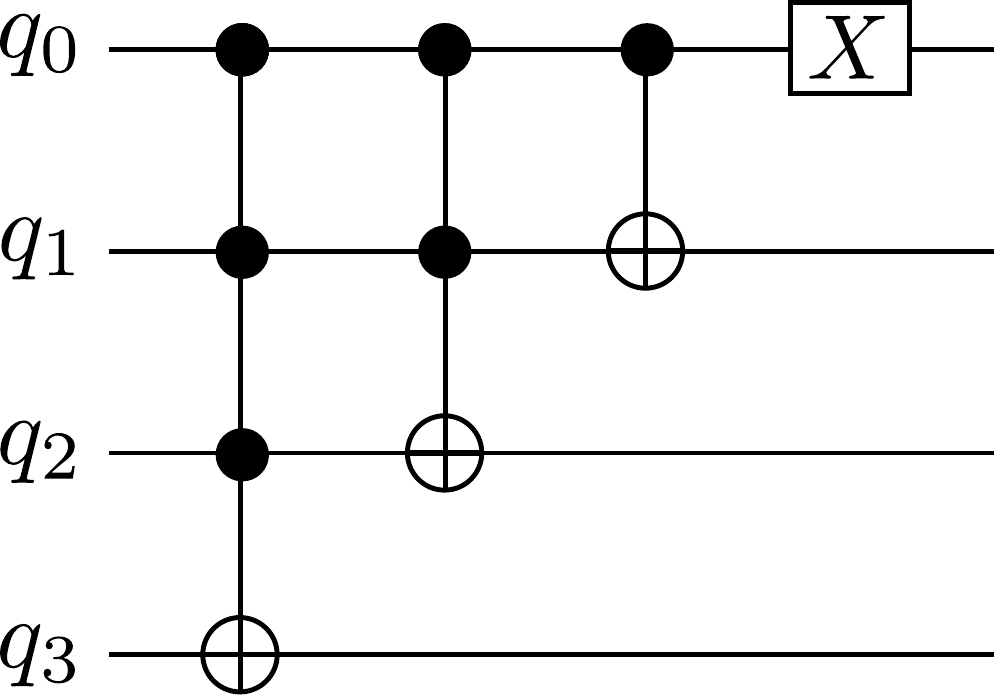}
\caption{Quantum circuit for the shift operator. \label{fig:shift}}
\end{figure}

The shift operator $P$ is represented by a sequence of multi-controlled Toffoli gates with at most $n-1$ control lines, as shown in Fig.~\ref{fig:shift}.
With $k \geq 3$ denoting the number of control lines, a $k$-controlled Toffoli gate can be decomposed into $2k-4$ relative-phase Toffoli gates and a Toffoli gate using $k-2$ auxiliary qubits \cite{Maslov2016advantages}.
Because a shift operator has $k$-controlled Toffoli gates for $3 \leq k \leq n-1$, a CNOT gate, and an X gate, it can be expressed by $(n-2)(n-3)$ relative-phase Toffoli gates, $n-3$ Toffoli gates, a CNOT gate, and an X gate.
Consequently, a quantum circuit for a shift operator requires $\mathcal{O}(n^2)$ depth and $2n-3$ qubits, including auxiliary qubits.
Note that this implementation of the shift operator $P$ is the major bottleneck for making a quantum circuit shallow, and so it is challenging to perform the proposed method directly on NISQ devices.
Thus, an efficient implementation of the shift operator is a crucial aspect that we will address in future research.
For example, Oomura et al. \cite{Oomura2021design} proposed an efficient implementation of the Toffoli gate employing the IBM Q Open Pulse Systems, which halves the gate time and improves the fidelity.
We believe that such pulse design approaches are promising.

The decomposition of the matrix $A$ introduced in a previous approach \cite{Liu2020variational} can also be used in our proposed method.
Though this decomposition yields $\mathcal{O}(n)$ terms to be measured, the fact that the shift operator is not used means that the quantum circuit remains shallow.
With the use of this decomposition, the previous method requires the expectations of $X \otimes A$ and $A^2$, which yield $2n+1$ and $4n+1$ terms to be measured, respectively, while our formulation requires the expectations of $X \otimes I$ and $A$, which yield $1$ and $2n+1$ terms to be measured, respectively.
Therefore, our proposed method has the advantage that the number of terms to be measured is roughly one-third of that in the previous method.

\subsection{Number of shots for expectation estimation} \label{sec:sampling}
To estimate the expectation value of a certain observable using a quantum computer, a quantum circuit with state preparation and measurement is run many times to sample an observable value and apply the Monte Carlo approach.
Each run of a quantum circuit to obtain a sample is referred to as a ``shot.''
In this subsection, we estimate the number of shots required to estimate the expectations.

In the proposed method, the expectations to be estimated are $\Braket{f, \psi(\bs{\theta}) | X \otimes I^{\otimes n} | f, \psi(\bs{\theta})}$ and $\Braket{\psi(\bs{\theta}) | A  | \psi(\bs{\theta})}$, where the latter consists of several terms depending on the boundary conditions, as mentioned in Sec.~\ref{sec:cost_eval}.
Let $q_i^{(j)}$ denote the $j$-th sample value for estimating the $i$-th expectation value.
Here, $q_1^{(j)}$ denotes the value of the $j$-th sample of  $\Braket{f, \psi(\bs{\theta}) | X \otimes I^{\otimes n} | f, \psi(\bs{\theta})}$, whereas $q_{\geq 2}^{(j)}$ denotes the value of the $j$-th sample of $\Braket{\psi(\bs{\theta}) | A  | \psi(\bs{\theta})}$.
In the following discussion, the parameter values $\bs{\theta}$ are assumed to be fixed and are omitted to simplify the notation.
Using $S_i$ shots, the $i$-th expectation value $\bar{q}_i$ is estimated as follows:
\begin{equation}
\bar{q}_i := \dfrac{ \sum_{j=1}^{S_i} q_i^{(j)}}{S_i}.
\end{equation}
Regarding $q_i^{(j)}$ as a random variable with a mean value of $\mu_i$ and a variance of $\sigma_i^2$, the mean and variance of $\bar{q}_i$ are written as
\begin{align}
\mathbb{E}[\bar{q}_i] &= \mu_i \\
\mathrm{Var}[\bar{q}_i] &= \dfrac{ \sum_{j=1}^{S_i} \mathrm{Var}[q_i^{(j)}] }{S_i^2} = \dfrac{ \sum_{j=1}^{S_i} \sigma_i^2 }{S_i^2} = \dfrac{ \sigma_i^2 }{S_i}.
\end{align}
The variance of $\bar{q}_i$ corresponds to a squared standard error of $q_i^{(j)}$, the standard error of which is $\sigma_i/\sqrt{S_i}$.

Now, the cost function $E_h$ in Eq.~(\ref{eq:E_h_theta}) is written using $\mu_i$, and is assumed to be estimated as follows using the approximated expectation values:
\begin{equation} 
E_h = -\dfrac{1}{2} \dfrac{\mu_1^2}{ \sum_{i=2}^m \mu_i }
     \approx -\dfrac{1}{2} \dfrac{\bar{q}_1^2}{ \sum_{i=2}^m \bar{q}_i } =: g(\bar{q}_1, \ldots, \bar{q}_m ), \label{eq:E_h_approx}
\end{equation}
with $g(\bar{q}_1, \ldots, \bar{q}_m )$ denoting the approximated cost function, where $m=3$ for the periodic boundary condition, $m=4$ for the Dirichlet boundary condition, and $m=5$ for the Neumann boundary condition, which corresponds to the number of terms to be evaluated derived in Sec.~\ref{sec:cost_eval}.

Note that sampling using quantum computers is performed independently for each term of the expectation, leading to $\mathrm{Cov}(\bar{q}_i, \bar{q}_{i^\prime})=0$ for $i \neq i^\prime$ where $\mathrm{Cov}(\bar{q}_i, \bar{q}_{i^\prime})$ is the covariance of $\bar{q}_i$ and $\bar{q}_{i^\prime}$. 
Using the first-order Taylor series expansion of $g(\bar{q}_1, \ldots, \bar{q}_m )$ around $\mu_i$ for $i \in [1, m]$, the mean squared error $\varepsilon^2$ between $E_h$ and $g(\bar{q}_1, \ldots, \bar{q}_m )$ in Eq.~(\ref{eq:E_h_approx}) is then written as follows:
\begin{align}
\varepsilon^2 = r_\mathrm{opt}^2 \left( \dfrac{\sigma_1^2}{S_1} + \dfrac{1}{4} r_\mathrm{opt}^2 \sum_{i=2}^m \dfrac{\sigma_i^2}{S_i} \right). \label{eq:mse}
\end{align}
A detailed derivation of the above equation is provided in Appendix~\ref{sec:mse}.
In the proposed method, for simplicity, we use the same number of shots, denoted by $S$, for all sampling processes, which yields
\begin{align}
\varepsilon^2 &\approx r_\mathrm{opt}^2 \left( \sigma_1^2 + \dfrac{1}{4} r_\mathrm{opt}^2 \sum_{i=2}^m \sigma_i^2 \right) \dfrac{1}{S}. \label{eq:mse_approx}
\end{align}
This equation implies that the mean square error is inversely proportional to the number of shots.

\subsection{Time complexity}
Here, we analyze the time complexity of the proposed method in terms of state preparation, number of quantum circuits, number of shots, and the number of iterations required to optimize the parameter set $\bs{\theta}$.
Note that the time complexity of the classical computing parts, such as parameter initialization and update, depends on the classical implementation and optimizers; thus, it is omitted from this analysis.

The time complexity of state preparation represents the time required for setting up the quantum circuit before performing the measurements to estimate a certain expectation value.
Therefore, the time complexity of state preparation, denoted by $T_P$, can be estimated by the depth of the quantum circuit of state preparation:
\begin{align}
T_P :=& ~ \mathcal{O}(D_\text{ansatz}+D_\text{enc}+D_\text{shift}) \\
  =& ~ \mathcal{O}(D_\text{ansatz}+D_\text{enc}+n^2)
\end{align}
where $D_\text{ansatz}$, $D_\text{enc}$, and $D_\text{shift}$ denote the circuit depths of the ansatz, amplitude encoding, and shift operator, respectively.

To estimate the cost function value using a quantum computer, several quantum circuits are required, corresponding to the numerator of Eq.~(\ref{eq:E_h_theta}) and each term in Eqs.~(\ref{eq:A_pauli_periodic})--(\ref{eq:A_pauli_N}) in the denominator of Eq.~(\ref{eq:E_h_theta}).
The required number of quantum circuits, $T_C$, is
\begin{equation}
T_C := \mathcal{O}(1),
\end{equation}
because $T_C=3$ for periodic boundary conditions, $T_C=4$ for Dirichlet boundary conditions, and $T_C=5$ for Neumann boundary conditions, independent of the scale of the problem, $n$.
Furthermore, when we use gradient-based optimizers, we need several quantum circuits, the number of which is proportional to the number of parameters in the parameter set $\bs{\theta}$, to evaluate the gradient of the cost function.
As the number of parameters is $\mathcal{O}(n D_\text{ansatz})$, the number of quantum circuits required, denoted by $T_G$, is
\begin{equation}
T_G := \mathcal{O}(n D_\text{ansatz}).
\end{equation}

To evaluate the cost function and its gradient, each quantum circuit must be run many times for the sampling required to estimate the expectation values.
Based on the discussion in Sec.~\ref{sec:sampling}, the required number of shots is
\begin{equation}
T_S := \mathcal{O} \left(\frac{1}{\varepsilon^2} \right).
\end{equation}

In the proposed method, the abovementioned evaluations of the cost function and its gradient through quantum circuits are repeated while the parameter set $\bs{\theta}$ is updated, as discussed in Sec.~\ref{sec:algo}.
The number of iterations is strongly dependent on the classical optimization solver and the terminal condition setting.
As the discussion of classical optimization solvers is beyond the scope of this paper, let $T_\text{it}$ denote the number of iterations, for simplicity.

Consequently, the total time complexity can be derived as
\begin{align}
T :=& ~ T_\text{it} T_P \left( T_C + T_G \right) T_S \nm \\
 =& ~ \mathcal{O} \left( \dfrac{ T_\text{it} \left( D_\text{ansatz}+D_\text{enc}+n^2 \right) n D_\text{ansatz} }{\varepsilon^2} \right). \label{eq:time_complexity}
\end{align}

The time complexity for solving the Poisson equation by classical computing is $\mathcal{O}(N \log N)$ \cite{Saad2003Iiterative}, where $N$ is the size of the matrix $A$, i.e., $N=2^n$. Thus, the proposed method has reduced time complexity compared with classical algorithms, as long as the number of optimization iterations $T_\text{it}$, the depth of the ansatz $D_\text{ansatz}$, and the depth of the quantum circuit for the amplitude encoding $D_\text{enc}$ are relatively small, i.e., $\mathcal{O} ( \text{poly}(n) )$.

The time complexity of the previous method \cite{Liu2020variational} can also be evaluated through a similar procedure.
Here, only the result is provided:
\begin{equation}
T = \mathcal{O}\left( \dfrac{T_\text{it}(D_\text{ansatz} + D_\text{enc})n^2 D_\text{ansatz}}{\varepsilon^2} \right).
\end{equation}
Hence, if either the depth of the ansatz or that of the amplitude encoding is greater than $\mathcal{O} ( n )$, the proposed method has reduced time complexity compared with the previous method.
As for amplitude encoding, a depth of $\mathcal{O} ( n^2 )$ is required to encode arbitrary real-valued inputs \cite{Araujo2021divide}, which implies that the proposed method has reduced time complexity compared with the previous method.

\subsection{Barren plateaus} \label{sec:plateau}
Note that the proposed cost function will suffer from the problem of exponentially vanishing gradients, i.e., barren plateaus, because of the definition of the cost function and the use of an alternating layered ansatz~\cite{Cerezo2021cost}.
Because the operator $A_{\mathcal{T}_\text{even}} = I^{\otimes {n-1}} \otimes (I-X)$ in the denominator of the cost function acts non-trivially on only one qubit, i.e., the operator is local, the expectation of the operator is resilient to the problem of barren plateaus.
However, the operator $A_{\mathcal{T}_\text{odd}} = P^{-1} \left( I^{\otimes {n-1}} \otimes (I-X) \right) P$ acts non-trivially on all qubits because of the shift operator, i.e., it is global, and so the expectation of the operator will be affected by barren plateaus.
Moreover, although the operator $X \otimes I$ in the numerator of the cost function is local, the expectation of the operator will be affected by barren plateaus because the numerator inherently evaluates the inner product of states $\Ket{\psi(\bs{\theta})}$ and $\Ket{f}$ by definition, which is a global quantity.
Consequently, the cost function as a whole will be affected by barren plateaus, even though one term is resilient to the problem.
Therefore, the required order of shots needed to evaluate the gradients will be exponential with respect to a certain error because of the exponentially vanishing gradients.
We will attempt to alleviate this problem in future research.
A more detailed discussion is provided in Appendix~\ref{sec:plateau_ex}.

\section{Numerical experiments} \label{sec:numel}
This section describes the results of several numerical experiments that demonstrate the validity of the proposed method.
The proposed method was implemented in Qiskit ver~0.23 \cite{Qiskit}, an open-source framework for working with quantum computers.
The \textit{statevector simulator} backend in Aer, which is a high-performance quantum computing simulator operating with Qiskit, was used for the calculations, except for those reported in Sec.~\ref{sec:ex_shots}, where the \textit{QASM simulator} backend was used.
As an optimizer for updating $\bs{\theta}$, we employed the Broyden--Fletcher--Goldfarb--Shanno method \cite{Broyden1970convergence, Fletcher1970new, Goldfarb1970family, Shanno1970conditioning}, with the gradient of the cost function evaluated by quantum computing.
An analytic derivation of the expression for the gradient is provided in Appendix~\ref{sec:grad}.

In the experiments reported below, the unitary $U_\text{b}$ for preparing $\Ket{f}$ was set as
\begin{equation}
    U_\text{b} = H^{\otimes n} X \otimes I^{\otimes (n-1)}, \label{eq:U_b}
\end{equation}
where $H$ represents a Hadamard gate.
This unitary $U_\text{b}$ prepares $\Ket{f}$ in the form of a step function from $1/2^{n/2}$ to $-1/2^{n/2}$.
The number of layers of the ansatz was fixed at $5$.

\subsection{Solutions for three types of boundary conditions}
\begin{figure}[t]
\centering
\subfloat[Periodic boundary conditions]{\includegraphics[width=8cm]{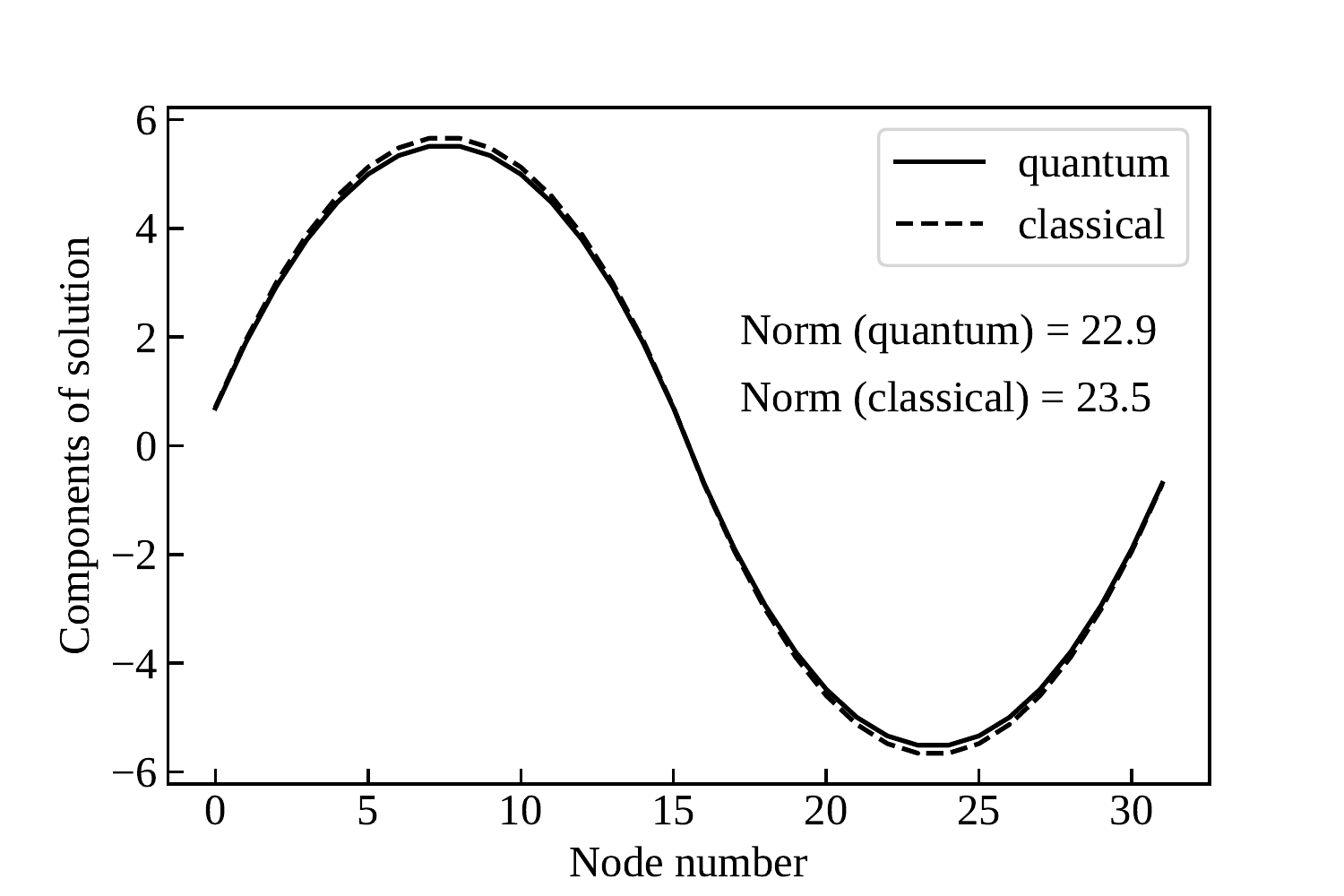}} \\
\subfloat[Dirichlet boundary conditions]{\includegraphics[width=8cm]{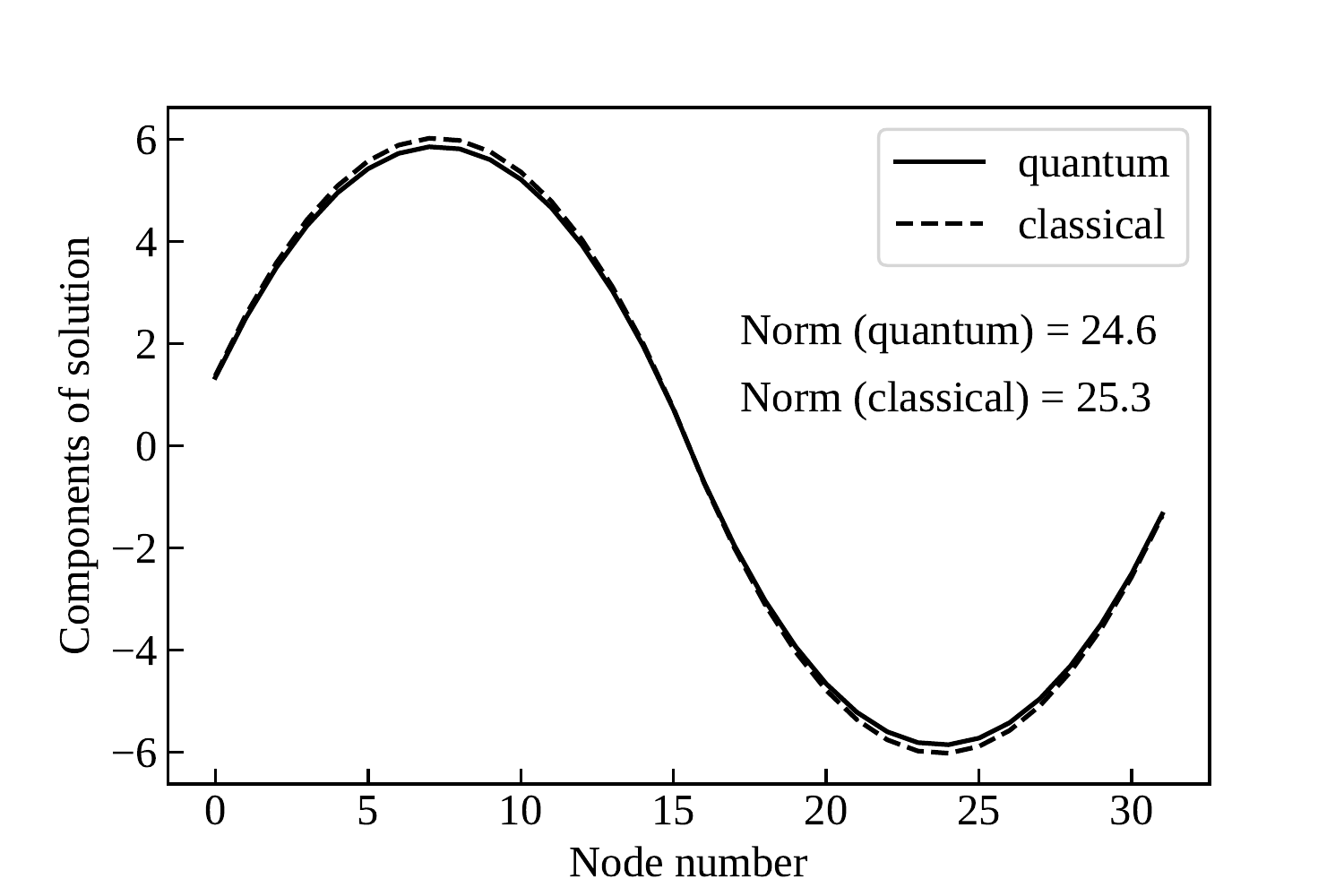}} \\
\subfloat[Neumann boundary conditions]{\includegraphics[width=8cm]{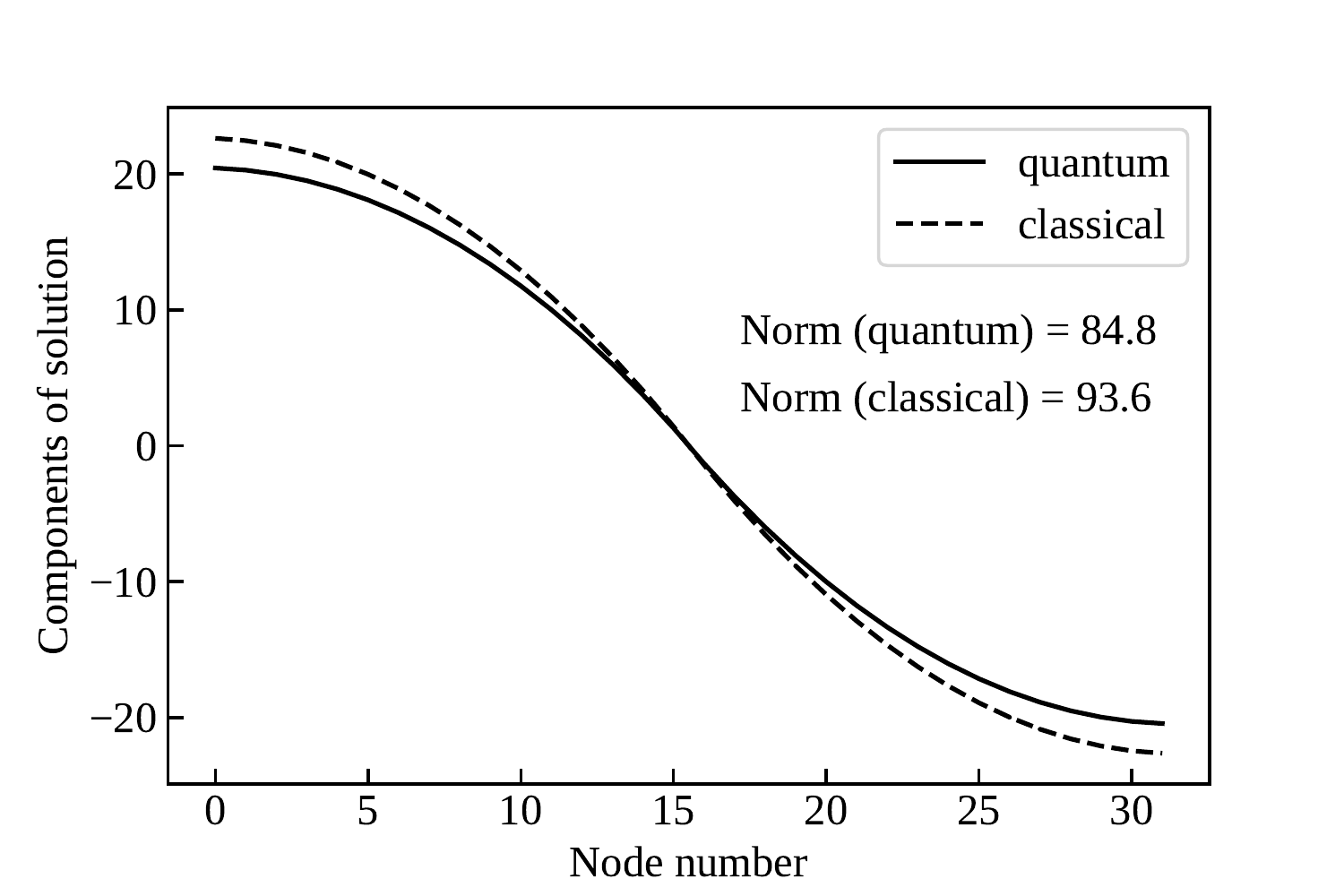}}
\caption{Distribution of solutions. The components of the solution vectors on each node are plotted. \label{fig:sol_field}}
\end{figure}
\begin{figure}
\includegraphics[width=8cm]{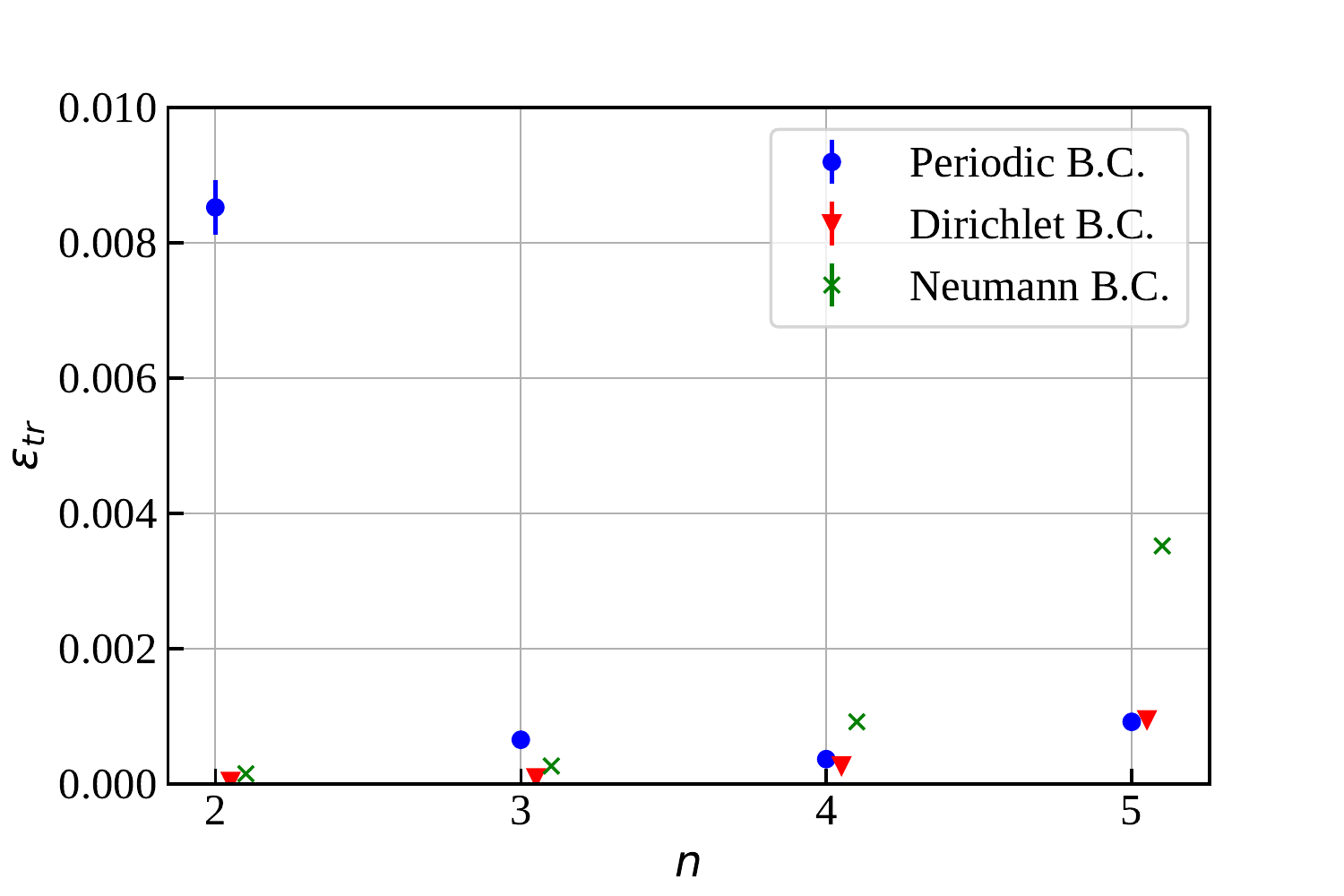}
\caption{Trace distances $\varepsilon_{tr}$ vs. the number of qubits $n$. The mean values of ten trials are plotted, with the error bars representing the standard deviations. 
The x-axis is slightly shifted for different legends for visibility.
\label{fig:trace_err}}
\end{figure}

First, we provide solutions obtained by the proposed method for the three types of boundary conditions to show the applicability of the proposed method to these basic types.
Here, the \textit{statevector simulator} backend in Aer was used to evaluate the proposed method in an ideal environment without any noise or sampling errors.

In this experiment, the optimization procedure was terminated when the norm of the gradients became less than the predetermined threshold value.
The optimization was performed ten times from randomly set initial parameters $\bs{\theta}$ for each condition.
When imposing the periodic or Neumann boundary conditions on both edges, a regularization term $\epsilon I$ with $\epsilon = 10^{-3}$ was added to the system matrix $A$ to prevent the matrix from becoming singular.

Figures~\ref{fig:sol_field}(a)--(c) show the components of the solution vectors on each node from ten trials when $n=5$.
As shown in these figures, the solutions for the periodic and Dirichlet boundary conditions are similar to each other, whereas the Neumann boundary condition gives a totally different solution when the input vector $\Ket{f}$ is prepared using the unitary operator in Eq.~(\ref{eq:U_b}).
Figures~\ref{fig:sol_field}(a)--(c) also indicate that the proposed algorithm underestimates the norms of the solution vectors, although the directions of the solution vectors given by the proposed method are in good agreement with those from classical computing.

Figure~\ref{fig:trace_err} shows the trace distance $\varepsilon_\text{tr}$ between the solutions obtained by the proposed method and the classical computation when calculating $A^{-1} \bs{f}$ with respect to the number of qubits.
The trace distance $\varepsilon_\text{tr}$ between the trial state $\Ket{\psi(\bs{\theta})}$ and the normalized ground-truth $\Ket{\bar{u}}:=\Ket{u}/ \sqrt{ \Braket{u|u} }$ is defined as
\begin{equation}
\varepsilon_\text{tr} := \dfrac{1}{2} \text{Tr} \left( \sqrt{\left( \Ket{\psi}\Bra{\psi} - \Ket{\bar{u}}\Bra{\bar{u}} \right)^2 } \right) = \sqrt{1 - \left|\Braket{\psi|\bar{u}} \right|^2}.
\end{equation}
The plots and error bars represent the mean values and the standard deviations of ten experiments.
This figure shows that the trace distance $\varepsilon_\text{tr}$ is less than $0.01$ for all cases, which results in the fidelity $\Braket{\psi|\bar{u}}^2$ being greater than $0.9999$.
These trace distance and fidelity values compare favorably with those reported in previous research \cite{Bravo2020variational, Liu2020variational}.
Therefore, the errors between the solution vectors obtained by the proposed algorithm and a classical approach are mainly caused by errors in the norms, and the directions of the solution vectors given by the proposed method actually agree with those from classical computing.
Because the errors of the solution depend on the Hilbert space expressed by the ansatz, we hope to conduct future research on the ansatz that is most suitable for expressing the solutions of certain PDEs.

\subsection{Comparison of the proposed method with the previous method}
We now compare the proposed method with the previous approach \cite{Liu2020variational} to evaluate the performance of the proposed method.
In the previous method, the cost function to be minimized is formulated as
\begin{align}
	E(\bs{\theta}) = \Braket{\psi(\bs{\theta}) | A \left( I - \Ket{f}\Bra{f} \right) A | \psi(\bs{\theta})}, \label{eq:Eh_previous}
\end{align}
which corresponds to the maximization of the cosine similarity of $A\Ket{\psi(\bs{\theta})}$ and $\Ket{f}$.
Because the cosine similarity does not take into account the norm information, the previous method does not explicitly provide the norm.
Note that the norm can be calculated once the normalized solution has been obtained.
Letting $r$ denote the norm of the solution, the equation $A r\Ket{ \psi(\bs{\theta})} = \Ket{f}$, which holds at the optimal point of $\bs{\theta}$ if the expressibility of the ansatz is sufficiently high, yields the following expression:
\begin{align}
    r = \dfrac{1}{\sqrt{\Braket{\psi(\bs{\theta}) | A^2 | \psi(\bs{\theta})}}}.
\end{align}
We also implemented the previous method using Qiskit by referring to the original paper.

In the following experiments, except those reported in Sec.~\ref{sec:ex_circ}, the proposed method used the Dirichlet boundary conditions imposed on both edges. 
The results are compared with those of the previous approach using the Dirichlet boundary conditions.

\subsubsection{Dependency of the number of circuit executions per cost function evaluation on the number of qubits} \label{sec:ex_circ}
First, we examined the dependency of the number of circuit executions on the number of qubits to evaluate the scalability of the proposed method.
For randomly set parameters $\bs{\theta} \in [0, 4\pi]$, the number of circuit executions for evaluating a cost function value was recorded for both the proposed and previous methods.

\begin{figure}[t]
\includegraphics[width=9cm]{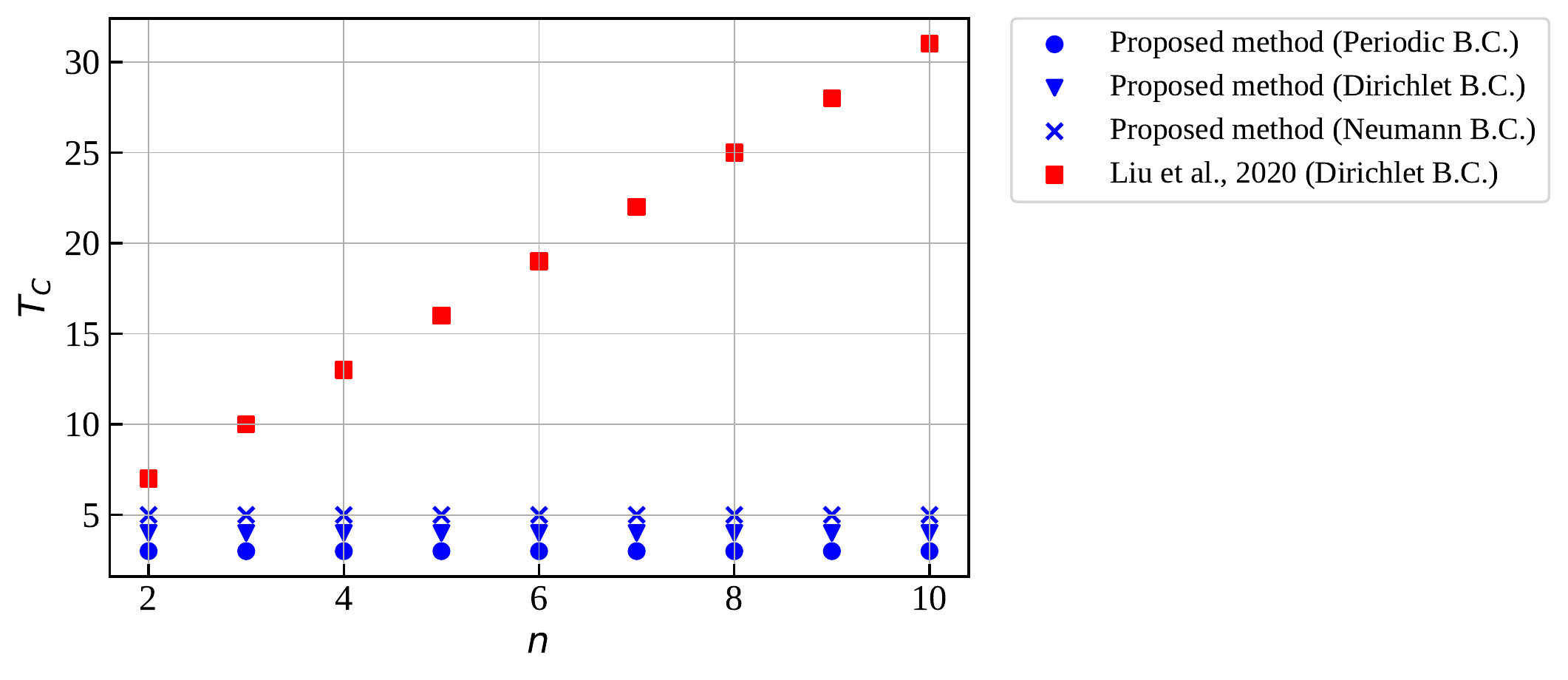}
\caption{Number of circuit executions per cost function evaluation $T_C$ vs. the number of qubits for both the proposed and previous methods \cite{Liu2020variational}. \label{fig:circ_count_comparison}}
\end{figure}

Figure~\ref{fig:circ_count_comparison} illustrates the number of circuit executions per cost function evaluation with respect to the number of qubits for both the proposed and previous methods \cite{Liu2020variational}.
This figure clearly shows that the proposed method only requires $\mathcal{O}(1)$ measurements per cost function evaluation, whereas the previous method requires $\mathcal{O}(n)$ measurements.
This result confirms the validity of the proposed formulation given in Sec.~\ref{sec:cost_eval}, implying that our proposed method significantly reduces the required number of expectation calculations on quantum computers.

\subsubsection{Dependency of the number of iterations on the number of qubits}
\begin{figure*}[t]
\includegraphics[width=0.9\textwidth]{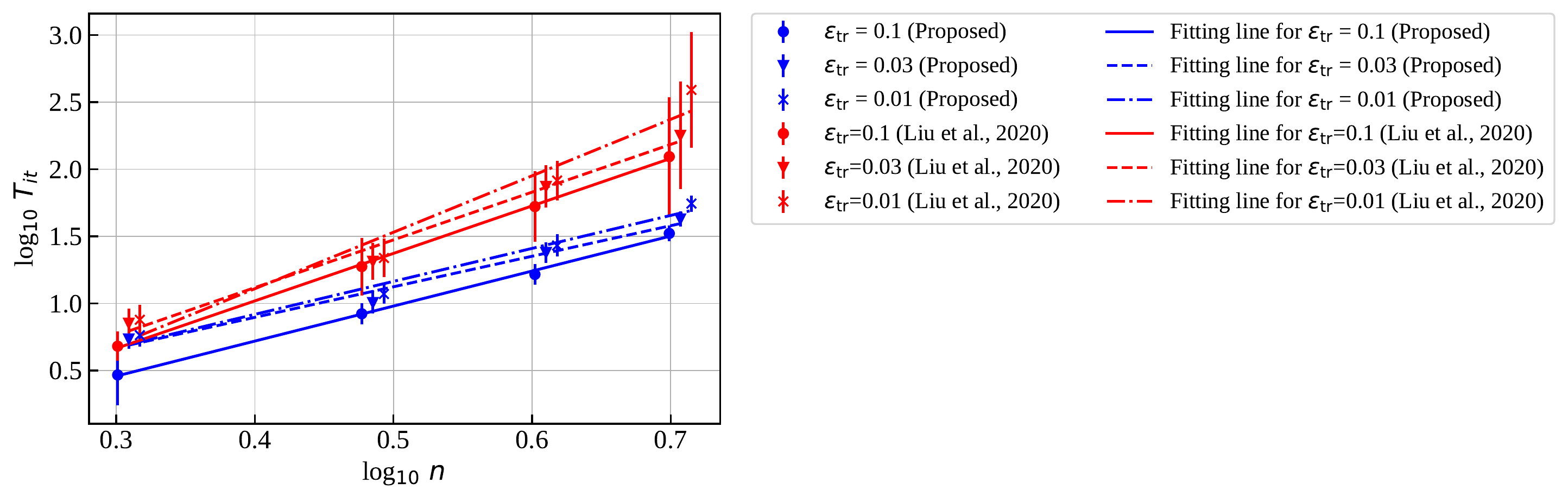}
\caption{Decadic logarithm of the number of optimization iterations $T_\mathrm{it}$ vs. the number of qubits $n$ for both the proposed and previous methods \cite{Liu2020variational}. The points show the mean values of ten experiments and the fitted lines minimize the squared errors of fitting. The error bars represent the standard deviation of ten experiments.
The x-axis is slightly shifted for different legends for visibility.
\label{fig:statesim_comparison}}
\end{figure*}

Next, we examined the dependency of the number of optimization iterations on the number of qubits to evaluate the scalability of the classical computing part of the proposed method.

In this experiment, the terminal condition was based on the tolerance of the trace distance $\varepsilon_\text{tr}$ between the trial state $\Ket{\psi(\bs{\theta})}$ and the normalized ground-truth $\Ket{\bar{u}}:=\Ket{u}/ \sqrt{ \Braket{u|u} }$, which is the same criterion used in previous research \cite{Bravo2020variational}.
Note that this metric only evaluates the difference in direction between the trial state and the true solution; the difference in the norm cannot be determined.
In spite of this defect, we used the metric to compare the proposed method with the previous approach according to the same criterion used in the previous method.
The tolerance of the trace distance was successively set to $\varepsilon_\text{tr} = 0.1, 0.03, 0.01$.

For each condition, the optimization was run ten times, using randomly set initial parameters in $[0, 4 \pi]$.
Figure~\ref{fig:statesim_comparison} shows the decadic logarithm of the number of optimization iterations with respect to the decadic logarithm of the number of qubits for both the proposed method and the previous method \cite{Liu2020variational}.
The plots and error bars represent the mean values and the standard deviations of ten experiments, respectively.
This figure clearly shows that the number of iterations is positively correlated with the number of qubits.
Although it is difficult to assert the time complexity of the number of iterations numerically because of the error bars, these plots were fitted to the lines that minimize the squared errors of fitting.
The lines for the proposed method have slopes of $2.6$ for $\varepsilon_\text{tr}=0.1$, $2.3$ for $\varepsilon_\text{tr}=0.03$, and $2.5$ for $\varepsilon_\text{tr}=0.01$.
This implies that $T_\text{it}$ in Eq.~(\ref{eq:time_complexity}) is at most $\mathcal{O}(n^{2.6})$ in this experiment.
For the previous method, the lines have slopes of $3.5$ for $\varepsilon_\text{tr}=0.1$, $3.6$ for $\varepsilon_\text{tr}=0.03$, and $4.2$ for $\varepsilon_\text{tr}=0.01$.
Therefore, it seems that the number of iterations with respect to the number of qubits is of similar order in both the proposed method and the previous method.

\begin{figure*}[t]
\includegraphics[width=0.8\textwidth]{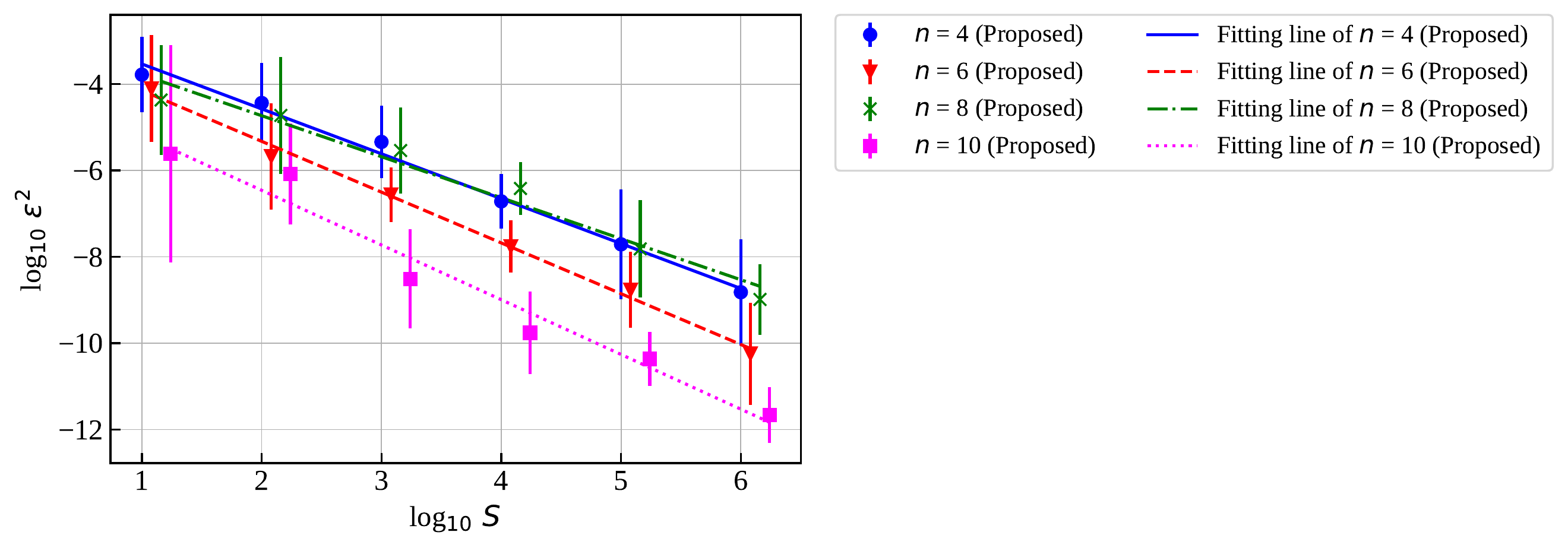}
\caption{Decadic logarithms of cost function error $\varepsilon$ vs. the number of shots $S$ for the proposed method. The optimization was run ten times with fixed parameters for each number of shots. The mean values are plotted, with the error bars representing the standard deviations. The mean and standard deviation of the slopes for $n=2, \ldots, 10$ (which includes cases that are not shown in this figure) are $-1.11$ and $0.12$, respectively.
The x-axis is slightly shifted for different legends for visibility.
\label{fig:objective_qasm_proposed}}
\end{figure*}
\begin{figure*}[t]
\includegraphics[width=0.8\textwidth]{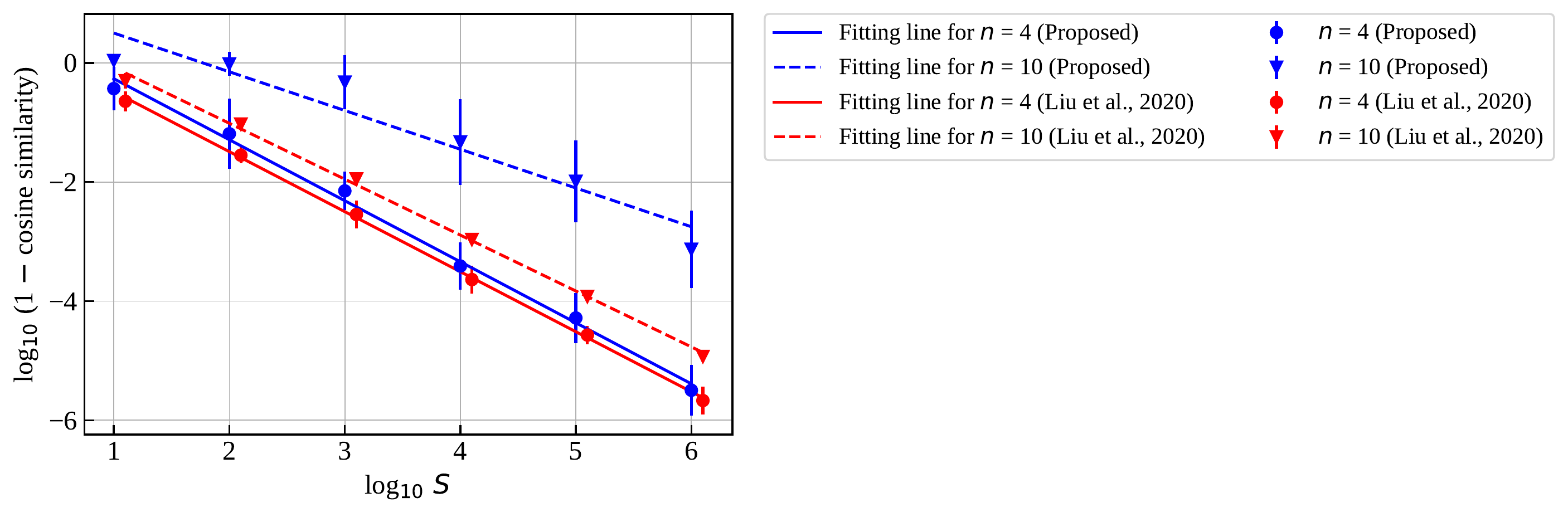}
\caption{Decadic logarithms of ``1 $-$ cosine similarity'' vs. the number of shots $S$ for the proposed and previous methods \cite{Liu2020variational} based on our implementation. The optimization was run ten times with fixed parameters for each number of shots. The mean values are plotted, with the error bars representing the standard deviations. The means and standard deviations of the slopes of the fitted lines for $n=2, \ldots, 10$ (which includes cases that are not shown in this figure) are $-0.99$ and $0.14$ for our proposed method and $-0.98$ and $0.03$ for the previous method, respectively.
The x-axis is slightly shifted for different legends for visibility.
\label{fig:grad_sim_comparison}}
\end{figure*}

\subsubsection{Dependency of the sampling errors on the number of shots} \label{sec:ex_shots}
Next, we examined the effect of the number of shots on the expectation estimations.
The \textit{QASM simulator} backend was used to evaluate the sampling errors in the environment without any noise.

For randomly set parameters $\bs{\theta} \in [0, 4\pi]$, the squared error between the ground-truth and the estimation value of the cost function was evaluated. 
The former was calculated by the \textit{statevector simulator} and the latter by the \textit{QASM simulator} (i.e., sampling).
The cost function was evaluated ten times by sampling for each number of shots.

Figure~\ref{fig:objective_qasm_proposed} shows the decadic logarithm of the squared error of the cost function with respect to the decadic logarithm of the number of shots for the proposed method.
The points show the mean values of the ten experiments and the error bars represent their standard deviations.
Fitting lines are also provided in the plots.
This figure clearly indicates that the squared error decreases as the number of shots increases.
The mean and standard deviation of the slopes are $-1.11$ and $0.12$, respectively, while the theoretical slope is $-1$, as derived in Eq.~(\ref{eq:mse_approx}).
Note that it does not make sense to compare the magnitude of errors between different numbers of qubits. Because this figure is plotted for a fixed parameter, which is set randomly for each number of qubits, the plots necessarily exhibit monotonicity with respect to $n$.
We also examined the effect of the number of shots on the expectation estimations for the previous method, and the results are illustrated in Appendix~\ref{sec:shoterror_ex}.

In addition to the cost function evaluation, it is important to precisely evaluate the gradients when using a gradient-based optimization method.
Therefore, we also evaluated the errors between the gradient evaluated by the sampling obtained using the \textit{QASM simulator} and that computed by the \textit{statevector simulator} with fixed parameters.

As a metric to evaluate the errors, we used the cosine similarity, which measures the similarity of directions, as the directions of the gradients are more important to optimizers than their norms.
Figure~\ref{fig:grad_sim_comparison} shows the decadic logarithm of ``1 $-$ cosine similarity'' with respect to the number of shots for both the proposed and previous methods.
This figure clearly shows that the cosine similarity increases as the number of shots increases.
When $n=10$, the slope of the fitting line increases.
This is caused by the existence of barren plateaus, whereby more shots are required to evaluate small gradients precisely.
The means and standard deviations of the slopes of the fitted lines for $n=2, \ldots, 10$ (which includes cases that do not appear in Fig.~\ref{fig:grad_sim_comparison}) are $-0.99$ and $0.14$ for our proposed method and $-0.98$ and $0.03$ for the previous method, respectively.
Therefore, it seems that the gradient estimation for a given number of shots is of similar order in both the proposed and previous methods.

However, we can clearly observe that the proposed method has longer error bars, i.e., a larger standard deviation of sampling, than the previous method.
This comes from the difference in the definition of the cost function, with the proposed method considering the norm of the solution vector as well as its direction in the optimization procedure.
This implies that more shots are required to estimate the norm of the solution in addition to its direction.

\section{Conclusions} \label{sec:conc}
This paper has presented a VQA for solving the Poisson equation based on the minimum potential energy.
The main contributions of the present study are as follows: 1) we have provided an explicit decomposition of the system matrix for the Poisson equation into $\mathcal{O}(1)$ terms consisting of simple observables, 2) the proposed method provides information about the norm of the solution vectors in addition to the direction of the vectors, and 3) the time complexity of the proposed algorithm has been derived and verified.
The first contribution implies that the proposed method only requires a small number of quantum measurements, compared with conventional approaches, at every iteration of the optimization procedure.
The second contribution enhances the ability of VQAs to solve PDEs, because the norm information is essential when using the calculation results for engineering developments.
As for the third contribution, we estimated the time complexity of the proposed method, and demonstrated that it has significant potential for reducing the computation time of classical computing algorithms.
The number of optimization iterations and the depth of the ansatz depend on the classical optimization and the architecture of the ansatz, respectively.
To derive the theoretical total time complexity, these aspects will be discussed in future work.

We believe the present study elevates the application of quantum computing to the field of computer-aided engineering and, moreover, design optimization.

\appendix
\section{Decomposition of stiffness matrix in finite element methods using graph theory} \label{sec:graph}
\begin{figure}[t]
\includegraphics[width=5cm]{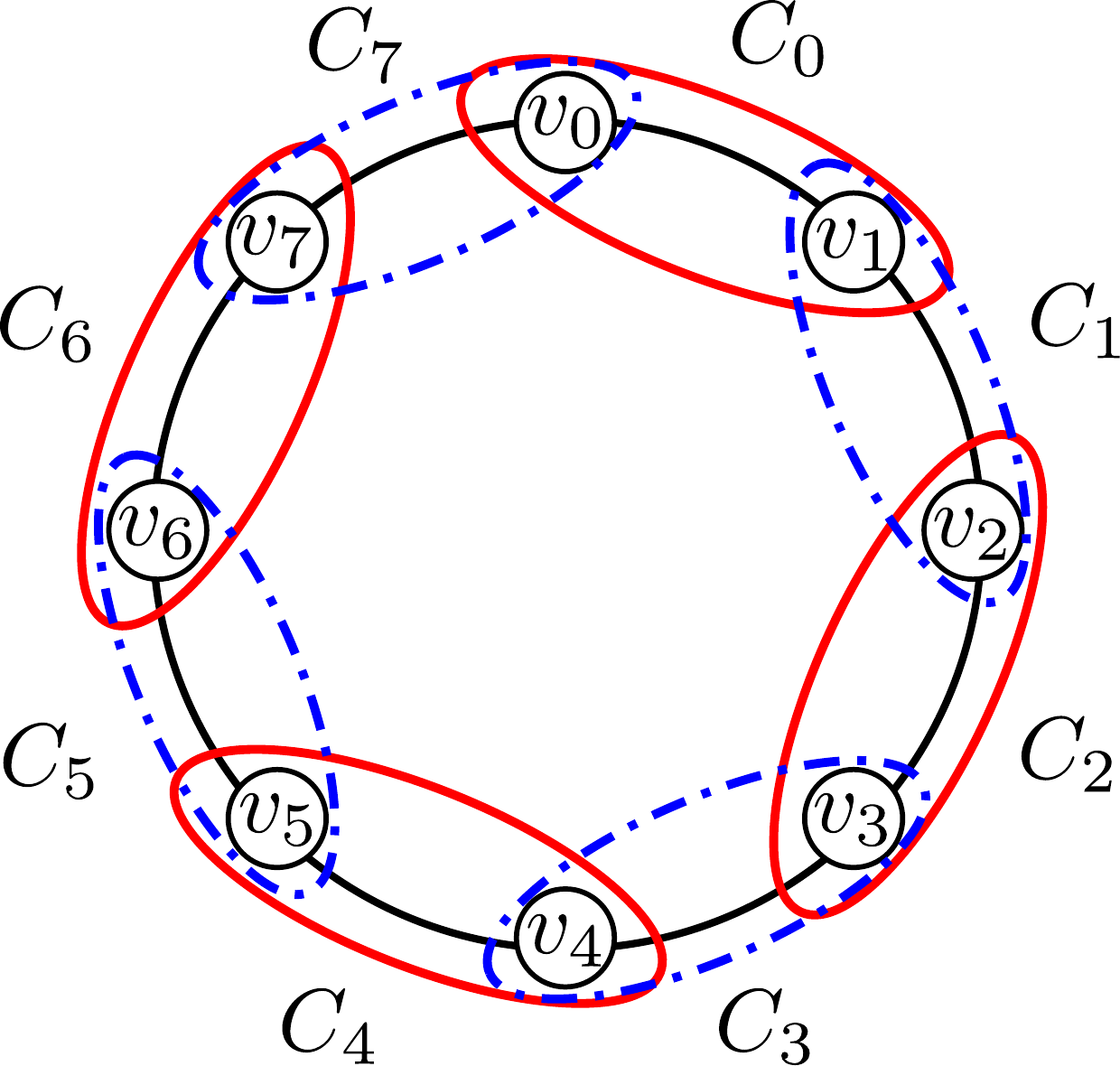}
\caption{Example of tessellations for an 8-node cycle graph. Red and blue ellipses indicate the tessellations $\mathcal{T}_\text{even}$ and $\mathcal{T}_\text{odd}$, respectively. \label{fig:tessellation}}
\end{figure}
First, a one-dimensional problem is considered.
Let us consider an $N$-node cycle graph $G=(V, E)$, where $V$ and $E$ represent node and edge sets, respectively.
This $N$-node cycle graph corresponds to the one-dimensional finite element with a periodic boundary condition.
A clique is defined as a subset of several nodes that form a complete subgraph.
A tessellation $\mathcal{T}$ is then defined as a set of cliques such that all nodes belong to one clique.
The tessellation $\mathcal{T}$ includes edges whose endpoints belong to the tessellation.
Generally, several tessellations can be defined in a graph, and there exists a set of tessellations such that all edges of the graph are included in at least one tessellation.
Such a set of tessellations is called a tessellation cover \cite{Abreu2018graph}.
Assuming that $N = 2^n$, two tessellations of the tessellation cover of the $N$-node cycle graph can be defined as
\begin{align}
\mathcal{T}_\text{even} &:= \{ C_{2i} ~|~ i \in [0, 2^{n-1}-1] \} \\
\mathcal{T}_\text{odd} &:= \{ C_{2i+1} ~|~ i \in [0, 2^{n-1}-1] \},
\end{align}
where $C_i := \{ v_i, v_{i+1} \}$ is a clique consisting of the $i$-th node $v_i$ and the $(i+1)$-th node $v_{i+1}$.
Note that we define $v_{N} := v_0$ to simplify the notation.

Figure~\ref{fig:tessellation} illustrates an example of these tessellations for an $8$-node cycle graph, where the red line represents a tessellation $\mathcal{T}_\text{even}$ and the blue dashed-dotted line represents the other tessellation $\mathcal{T}_\text{odd}$.

The decomposed matrices $A_{\mathcal{T}_\text{even}}$ and $A_{\mathcal{T}_\text{odd}}$ in Eqs.~(\ref{eq:A_even}) and (\ref{eq:A_odd}) can then be expressed as the sums of the element stiffness matrices related to elements in each tessellation $\mathcal{T}_\text{even}$ and $\mathcal{T}_\text{odd}$, respectively.

\begin{figure*}[t]
\centering
\subfloat[An element]{\includegraphics[width=1.5cm]{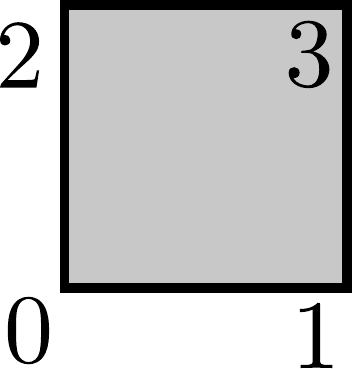}\label{fig:one_element}} \qquad
\subfloat[Two-dimensional mesh]{\includegraphics[width=4cm]{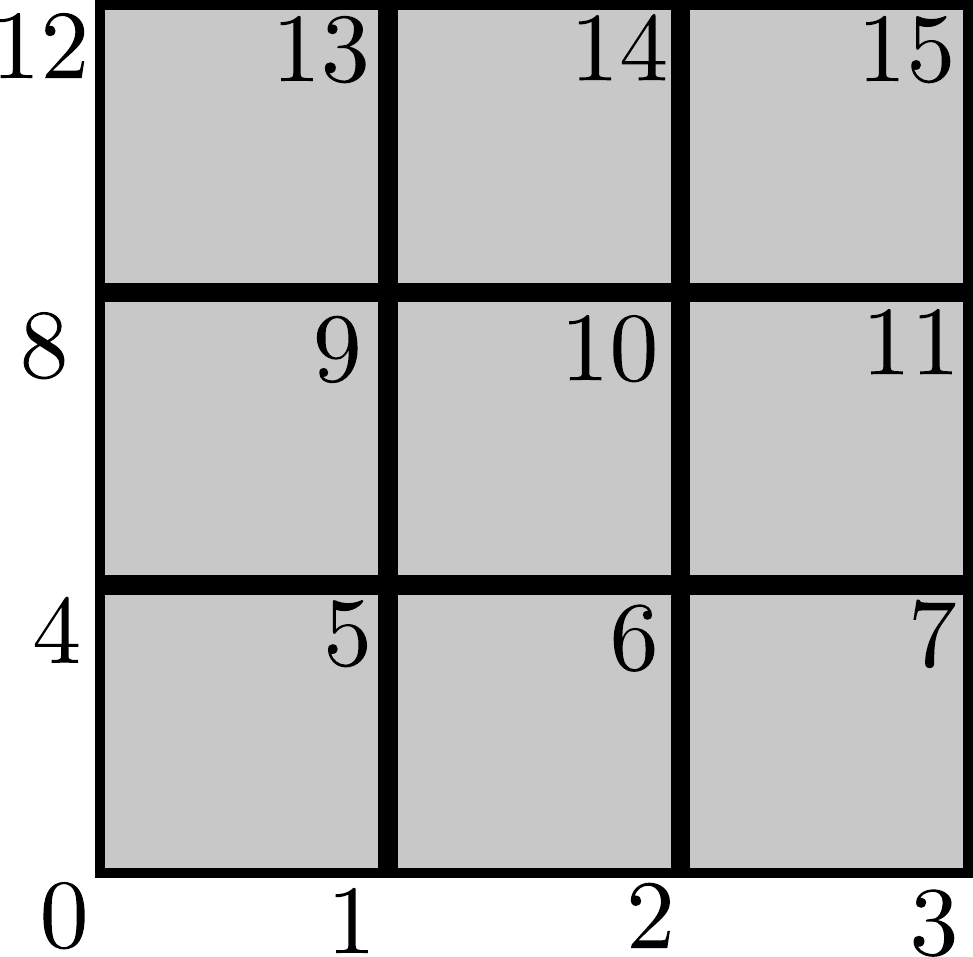}\label{fig:mesh_2d}} \qquad
\subfloat[Graph for two-dimensional mesh]{\includegraphics[width=6cm]{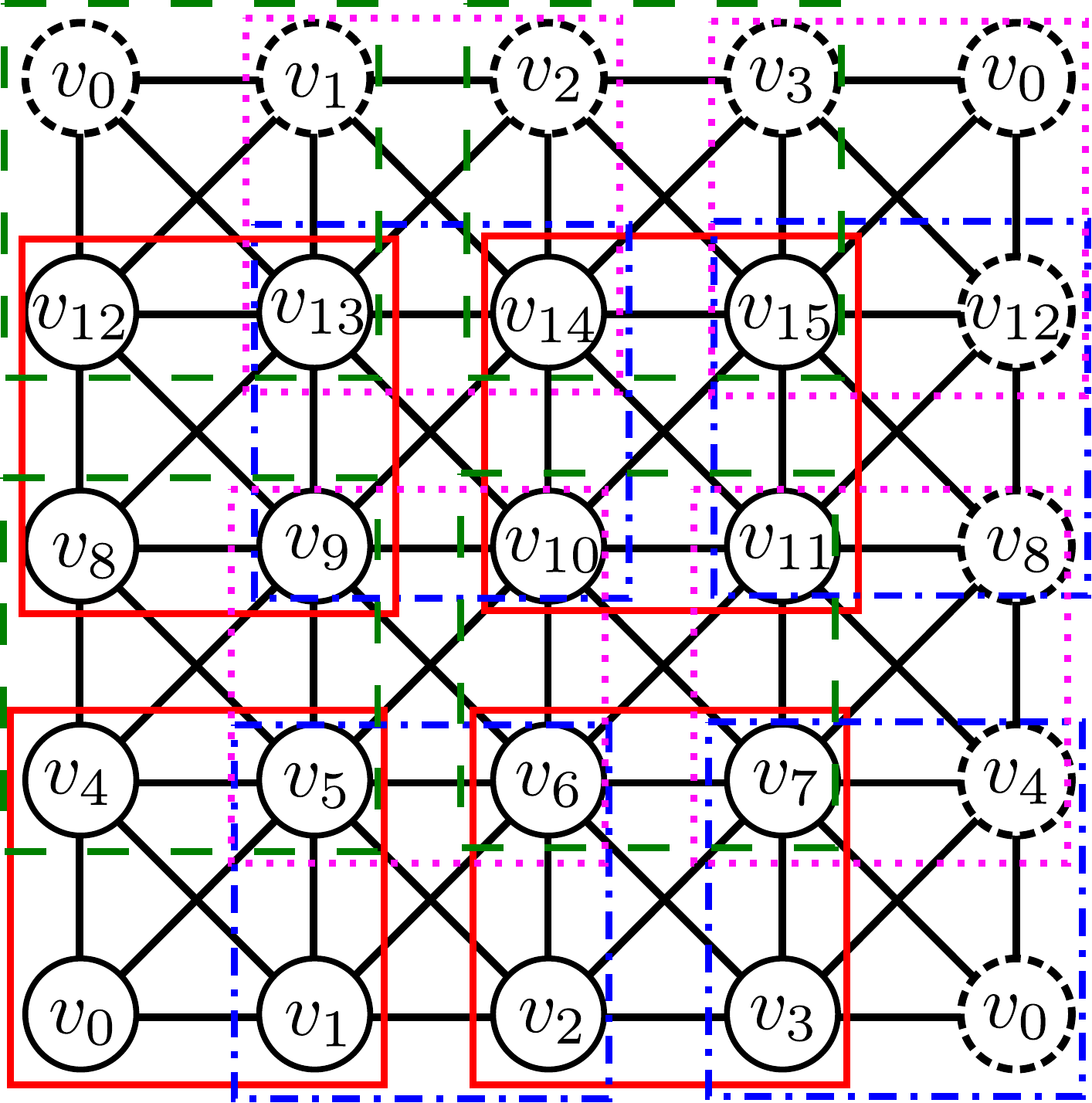}\label{fig:graph_for_2d}}
\caption[]{Example of two-dimensional finite elements and the corresponding graph: \subref{fig:one_element} an element with node numbers, \subref{fig:mesh_2d} a $3 \times 3$ two-dimensional square mesh, \subref{fig:graph_for_2d} a graph corresponding to the two-dimensional mesh. Rectangles drawn by red lines, blue dashed-dotted lines, green dashed lines, and magenta dotted lines indicate separate tessellations. \label{fig:tessellation_2d}}
\end{figure*}

The above discussion can easily be extended to two-dimensional problems.
Let us consider the finite element method in the two-dimensional Poisson equation.
For a first-order quadrilateral element of length $1$ in Fig.~\ref{fig:tessellation_2d}\subref{fig:one_element}, the element stiffness matrix is described as
\begin{align}
A_\text{e} &:= \dfrac{1}{6}
\begin{bmatrix}
4 & -1 & -1 & -2 \\
-1 & 4 & -2 & -1 \\
-1 & -2 & 4 & -1 \\
-2 & -1 & -1 & 4 \\
\end{bmatrix} \nm \\
&= \dfrac{1}{6} \left( 4I \otimes I - I \otimes X - X \otimes I - 2X \otimes X \right). \label{eq:A_e}
\end{align}
Now, let us describe the total stiffness matrix for the mesh in Fig.~\ref{fig:tessellation_2d}\subref{fig:mesh_2d} using the Pauli operators.
Here, for simplicity, we assume that periodic boundary conditions are imposed on all edges of the mesh.
Let $N_x$ and $N_y$ denote the numbers of columns and rows of nodes, respectively.
That is, the number of nodes is $N= N_x \times N_y$.
For example, in Fig.~\ref{fig:tessellation_2d}\subref{fig:mesh_2d}, $N_x=4$ and $N_y=4$.

We now define a graph corresponding to the mesh, as shown in Fig.~\ref{fig:tessellation_2d}\subref{fig:graph_for_2d}.
Each node $v_i$ of the graph corresponds to node $i$ of the mesh, and the graph has edges between nodes within the same elements.
Because of the periodic boundary conditions, nodes corresponding to the edges of the mesh are also connected, e.g., $v_0$ and $v_3$, $v_0$ and $v_{12}$, and so on.
For clear visibility, nodes with dashed circles are added on the upper and right sides.

Assuming that $N_x = 2^{n_x}$ and $N_y = 2^{n_y}$, four tessellations in the tessellation cover of the $N$-node graph can be defined as
\begin{widetext}
\begin{align}
\mathcal{T}_\text{0} &:= \{ C_{2i_x+2i_y N_x} ~|~ i_x \in [0, 2^{n_x-1}-1], i_y \in [0, 2^{n_y-1}-1] \} \\
\mathcal{T}_\text{1} &:= \{ C_{2i_x+1+2i_y N_x} ~|~ i_x \in [0, 2^{n_x-1}-1], i_y \in [0, 2^{n_y-1}-1] \} \\
\mathcal{T}_\text{2} &:= \{ C_{2i_x+(2i_y+1) N_x} ~|~ i_x \in [0, 2^{n_x-1}-1], i_y \in [0, 2^{n_y-1}-1] \} \\
\mathcal{T}_\text{3} &:= \{ C_{2i_x+1+(2i_y+1) N_x} ~|~ i_x \in [0, 2^{n_x-1}-1], i_y \in [0, 2^{n_y-1}-1] \},
\end{align}
\end{widetext}
where $C_i := \{ v_i, v_{i+1}, v_{N_x}, v_{N_x+1} \}$ is the $i$-th clique defined on the graph.
Note that we define $v_{i \geq N} := v_{i~\text{mod}~N}$ to simplify the notation.
In Fig.~\ref{fig:tessellation_2d}\subref{fig:graph_for_2d}, rectangles with red lines, blue dashed-dotted lines, green dashed lines, and magenta dotted lines represent $\mathcal{T}_\text{0}$, $\mathcal{T}_\text{1}$, $\mathcal{T}_\text{2}$, and $\mathcal{T}_\text{3}$, respectively.

Let $\Ket{i}$ be the quantum state corresponding to node $v_i$ of the graph. 
The quantum state $\Ket{i}$ consists of two quantum registers, $\Ket{i_x}$ and $\Ket{i_y}$, for each direction, as $\Ket{i} := \Ket{i_y} \Ket{i_x}$,
where $\Ket{i_x}$ and $\Ket{i_y}$ consist of $n_x$ and $n_y$ qubits, respectively.
The sum of the element stiffness matrix for the elements related to the tessellation $\mathcal{T}_{0}$, denoted as $A_{\mathcal{T}_{0}}$, can then be expressed as
\begin{widetext}
\begin{align}
A_{\mathcal{T}_{0}} =& \dfrac{1}{6} \left( 4 (I^{\otimes n_y-1} \otimes I) \otimes (I^{\otimes n_x-1} \otimes I) - (I^{\otimes n_y-1} \otimes I) \otimes (I^{\otimes n_x-1} \otimes X) \right. \nm \\
&- \left. (I^{\otimes n_y-1} \otimes X) \otimes (I^{\otimes n_x-1} \otimes I) - 2 (I^{\otimes n_y-1} \otimes X) \otimes (I^{\otimes n_x-1} \otimes X) \right) \nm \\
=& \dfrac{1}{6} \left( 4 I^{\otimes (n_x+n_y)} - I^{\otimes n_y} \otimes (I^{\otimes n_x-1} \otimes X) - (I^{\otimes n_y-1} \otimes X) \otimes I^{\otimes n_x} - 2 (I^{\otimes n_y-1} \otimes X) \otimes (I^{\otimes n_x-1} \otimes X) \right), \label{eq:A_0}
\end{align}
\end{widetext}
where the first $n_y$ tensor products are for the $y$-direction (row direction) and the latter $n_x$ tensor products are for the $x$-direction (column direction).
As the nodes of the cliques in the tessellation $\mathcal{T}_{1}$ can be expressed by adding $1$ to the node numbers of the nodes of cliques in the tessellation $\mathcal{T}_{0}$, the sum of the element stiffness matrix for the elements related to the tessellation $\mathcal{T}_{1}$, denoted as $A_{\mathcal{T}_{1}}$, can be described as follows:
\begin{equation}
A_{\mathcal{T}_{1}} = P_x^{-1} A_{\mathcal{T}_{0}} P_x,
\end{equation}
where $P_x$ is a shift operator in the $x$-direction defined as
\begin{equation}
P_x := \sum_{\substack{i_x \in [0, 2^{n_x}-1] \\ i_y \in [0, 2^{n_y}-1]}} \Ket{i_y}\Ket{(i_x+1)~\mathrm{mod}~2^{n_x}} \Bra{i_y}\Bra{i_x}.
\end{equation}
Similarly, the sum of the element stiffness matrix for the elements related to the tessellations $\mathcal{T}_{2}$ and $\mathcal{T}_{3}$, denoted as $A_{\mathcal{T}_{2}}$ and $A_{\mathcal{T}_{3}}$, respectively, can be described as
\begin{align}
A_{\mathcal{T}_{2}} &= P_y^{-1} A_{\mathcal{T}_{0}} P_y,\nm \\
A_{\mathcal{T}_{3}} &= P_x^{-1} P_y^{-1} A_{\mathcal{T}_{0}} P_x P_y
\end{align}
where $P_y$ is a shift operator in the $y$-direction defined as
\begin{equation}
P_y := \sum_{\substack{i_x \in [0, 2^{n_x}-1] \\ i_y \in [0, 2^{n_y}-1]}} \Ket{(i_y+1)~\mathrm{mod}~2^{n_y}}\Ket{i_x} \Bra{i_y}\Bra{i_x}.
\end{equation}
Consequently, the total stiffness matrix $\bs{A}$ can be described as the sum of the stiffness matrices related to each tessellation:
\begin{equation}
A = A_{\mathcal{T}_{0}} + A_{\mathcal{T}_{1}} + A_{\mathcal{T}_{2}} + A_{\mathcal{T}_{3}}.
\end{equation}
For Dirichlet and Neumann boundary conditions, we just have to add terms to adjust the stiffness matrices of edge elements.

\section{Derivation of the mean squared error between the exact cost function and that estimated by sampling} \label{sec:mse}
Here, the mean squared error between the exact cost function value and that estimated by sampling is derived using the Taylor series expansion.
The first-order Taylor series expansion of $g(\bar{q}_1, \ldots, \bar{q}_m )$ around $\mu_i$ for $i \in [1, m]$ is given as
\begin{align}
&g(\bar{q}_1, \ldots, \bar{q}_m) \nm \\
&= E_h + \sum_{i=1}^m \left. \pdif{g}{\bar{q}_i} \right|_{\bar{q}_i=\mu_i} (\bar{q}_i - \mu_i) + o \left( (\bar{q}_i - \mu_i)^2 \right),
\end{align}
where
\begin{equation}
g(\bar{q}_1, \ldots, \bar{q}_m) = -\dfrac{1}{2} \dfrac{\bar{q}_1^2}{ \sum_{i=2}^m \bar{q}_i }.
\end{equation}
Assuming that $\mathrm{Cov}(\bar{q}_i, \bar{q}_{i^\prime})=0$ for $i \neq i^\prime$, the mean squared error between the exact cost function value and that estimated by sampling can be evaluated as follows:
\begin{widetext}
\begin{align}
\varepsilon^2 &= \mathbb{E}[(g - E_h)^2] \nm \\
&\approx \mathbb{E} \left[ \left( \sum_{i=1}^m \left. \pdif{g}{\bar{q}_i} \right|_{\bar{q}_i=\mu_i} (\bar{q}_i - \mu_i) \right)^2 \right] \nm \\
&= \sum_{i=1}^m \sum_{j=1}^m \left. \pdif{g}{\bar{q}_i} \right|_{\bar{q}_i=\mu_i} \left. \pdif{g}{\bar{q}_j} \right|_{\bar{q}_j=\mu_j} \mathbb{E} \left[(\bar{q}_i - \mu_i) (\bar{q}_j - \mu_j) \right] \nm \\
&= \sum_{i=1}^m \left( \left. \pdif{g}{\bar{q}_i} \right|_{\bar{q}_i=\mu_i}  \right)^2 \dfrac{\sigma_i^2}{S_i} \quad (\because \mathrm{Cov}(\bar{q}_i, \bar{q}_j)=0)\nm \\
&= \dfrac{\mu_1^2}{\left( \sum_{i=2}^m \mu_i \right)^2} \dfrac{\sigma_1^2}{S_1} + \dfrac{1}{4}\dfrac{\mu_1^4}{ \left( \sum_{i=2}^m \mu_i \right)^4 } \sum_{i=2}^m \dfrac{\sigma_i^2}{S_i} \nm \\
&= \dfrac{\mu_1^2}{\left( \sum_{i=2}^m \mu_i \right)^2} \left( \dfrac{\sigma_1^2}{S_1} + \dfrac{1}{4} \dfrac{\mu_1^2}{ \left( \sum_{i=2}^m \mu_i \right)^2 } \sum_{i=2}^m \dfrac{\sigma_i^2}{S_i} \right) \nm \\
&= r_\mathrm{opt}^2 \left( \dfrac{\sigma_1^2}{S_1} + \dfrac{1}{4} r_\mathrm{opt}^2 \sum_{i=2}^m \dfrac{\sigma_i^2}{S_i} \right),
\end{align}
\end{widetext}
where $\delta_{ij}$ is Kronecker's delta.
The assumption that $\mathrm{Cov}(\bar{q}_i, \bar{q}_{i^\prime})=0$ for $i \neq i^\prime$ is based on the assumption that, in quantum computers, each shot is mutually independent.
In the last transformation, we have used the following equation:
\begin{align}
r_\mathrm{opt} &= \dfrac{ \Braket{f, \psi | X \otimes I^{\otimes n} | f, \psi} }{ \Braket{\psi | A | \psi} } \nm \\
&= \dfrac{\mu_1}{ \sum_{i=2}^m \mu_i }.
\end{align}

\section{Derivative of the cost function} \label{sec:grad}
The gradient of the cost function in Eq.~(\ref{eq:E_h_theta}) is now derived.
The partial derivative of the cost function with respect to the parameters $\bs{\theta}$ yields
\begin{widetext}
\begin{align}
    \pdif{E_h}{\bs{\theta}} &= - \dfrac{ \left( \Braket{f, \psi(\bs{\theta}) | X \otimes I^{\otimes n} | f, \psi(\bs{\theta})} \right) \pdif{ }{\bs{\theta}} \Braket{f, \psi(\bs{\theta}) | X \otimes I^{\otimes n} | f, \psi(\bs{\theta})} }{ \Braket{\psi(\bs{\theta}) | A  | \psi(\bs{\theta})} } \nm \\
    & \quad + \dfrac{1}{2} \dfrac{ \left( \Braket{f, \psi(\bs{\theta}) | X \otimes I^{\otimes n} | f, \psi(\bs{\theta})} \right)^2 \pdif{}{\bs{\theta}} \Braket{\psi(\bs{\theta}) | A  | \psi(\bs{\theta})} }{ \Braket{\psi(\bs{\theta}) | A  | \psi(\bs{\theta})}^2 } \label{eq:dEdth}.
\end{align}
\end{widetext}
Recalling that $\Ket{\psi (\bs{\theta})} = U(\bs{\theta}) \Ket{0}^{\otimes n}$, where $U(\bs{\theta})$ is a sequence of parameterized quantum gates, the following holds for the $i$-th parameter $\theta_i$:
\begin{align}
    \pdif{}{\theta_i} \Ket{\psi(\bs{\theta})} &= \pdif{}{\theta_i} U(\bs{\theta}) \Ket{0}^{\otimes n} \nm \\
    &= \dfrac{1}{2} U(\theta_1, \theta_2, \ldots, \theta_i+\pi, \ldots) \Ket{0}^{\otimes n},
\end{align}
under the assumption that the parameterized gates consist of either $R_X$, $R_Y$, or $R_Z$ gates.
Note that 
\begin{equation}
\Ket{\psi(\bs{\theta})_{,i}} := U(\theta_1, \theta_2, \ldots, \theta_i+\pi, \ldots) \Ket{0}^{\otimes n}
\end{equation}
is a quantum state because $U(\theta_1, \theta_2, \ldots, \theta_i+\pi, \ldots)$ is a unitary operator.

Now, recall that $\Ket{f, \psi(\bs{\theta})} := \left( \Ket{0}\Ket{f} + \Ket{1}\Ket{\psi(\bs{\theta})} \right) / \sqrt{2}$, which yields
\begin{align}
&\pdif{ }{\theta_i} \Braket{f, \psi(\bs{\theta}) | X \otimes I^{\otimes n} | f, \psi(\bs{\theta})} \nm \\
&= \dfrac{1}{4} \left( \Braket{\psi(\bs{\theta})_{,i}  | f } + \Braket{f | \psi(\bs{\theta})_{,i}} \right) \nm \\
&= \dfrac{1}{2} \Braket{f, \psi(\bs{\theta})_{,i} | X \otimes I^{\otimes n} | f, \psi(\bs{\theta})_{,i}}, \label{eq:dXdth}
\end{align}
where $\Ket{f, \psi(\bs{\theta})_{,i}} := \left( \Ket{0}\Ket{f} + \Ket{1}\Ket{\psi(\bs{\theta})_{,i}} \right) / \sqrt{2}$.
The following equation also holds:
\begin{align}
    &\pdif{ }{\theta_i} \Braket{\psi(\bs{\theta}) | A  | \psi(\bs{\theta})} \nm \\
    &= \dfrac{1}{2} \left(  \Braket{\psi(\bs{\theta})_{,i} | A  | \psi(\bs{\theta})} + \Braket{\psi(\bs{\theta}) | A  | \psi(\bs{\theta})_{,i} } \right) \nm \\
    &= \Braket{\psi(\bs{\theta})_{,i}, \psi(\bs{\theta}) | X \otimes A | \psi(\bs{\theta})_{,i}, \psi(\bs{\theta})}, \label{eq:dAdth}
\end{align}
where $\Ket{\psi(\bs{\theta})_{,i}, \psi(\bs{\theta})} := \left( \Ket{0}\Ket{\psi(\bs{\theta})_{,i}} + \Ket{1}\Ket{\psi(\bs{\theta})} \right) / \sqrt{2}$.

Substituting Eqs.~(\ref{eq:dXdth}) and (\ref{eq:dAdth}) into Eq.~(\ref{eq:dEdth}), the gradient of the cost function is derived as 
\begin{widetext}
\begin{align}
    \pdif{E_h}{\theta_i} &= - \dfrac{1}{2} \dfrac{ \left( \Braket{f, \psi(\bs{\theta}) | X \otimes I^{\otimes n} | f, \psi(\bs{\theta})} \right) \Braket{f, \psi(\bs{\theta})_{,i} | X \otimes I^{\otimes n} | f, \psi(\bs{\theta})_{,i}} }{ \Braket{\psi(\bs{\theta}) | A  | \psi(\bs{\theta})} } \nm \\
    & \quad + \dfrac{1}{2} \dfrac{ \left( \Braket{f, \psi(\bs{\theta}) | X \otimes I^{\otimes n} | f, \psi(\bs{\theta})} \right)^2 \Braket{\psi(\bs{\theta})_{,i}, \psi(\bs{\theta}) | X \otimes A | \psi(\bs{\theta})_{,i}, \psi(\bs{\theta})} }{ \Braket{\psi(\bs{\theta}) | A  | \psi(\bs{\theta})}^2 }.
\end{align}
\end{widetext}
Now, as all components of the gradient can be described as the expectations of observables, the gradient can be evaluated by quantum computers.

\section{Supplementary results of numerical experiments} \label{sec:result_ex}

\subsection{Barren plateaus} \label{sec:plateau_ex}
%%%%%%%%%%%%%%%%%%%%%%%%
\begin{figure*}[t]
\centering
\subfloat[Norm of $\partial E_h / \partial \boldsymbol{\theta}$]{\includegraphics[width=7cm]{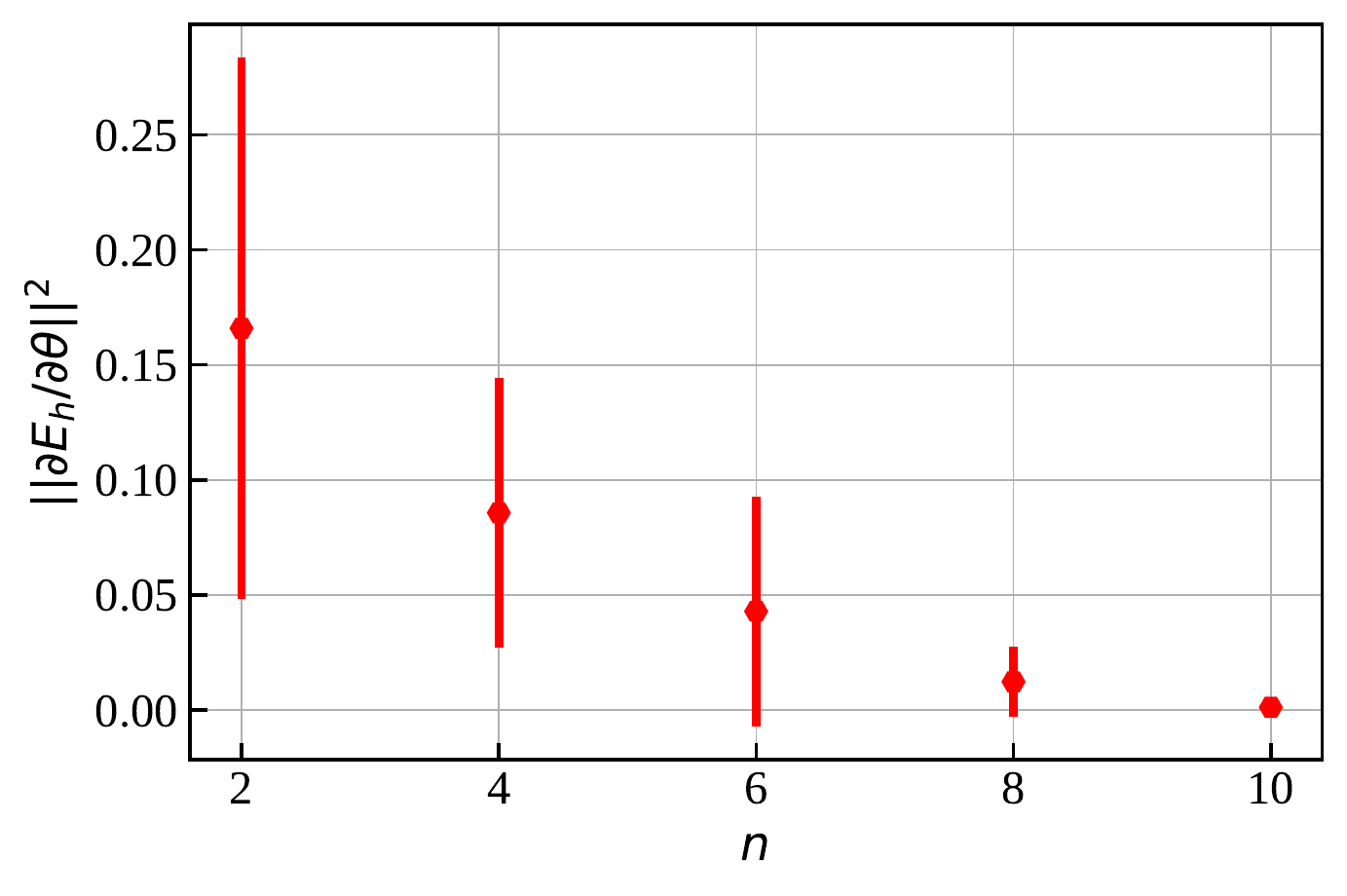}}
\subfloat[Norm of $\partial \Braket{A_\text{even}} / \partial \boldsymbol{\theta}$ ]{\includegraphics[width=7cm]{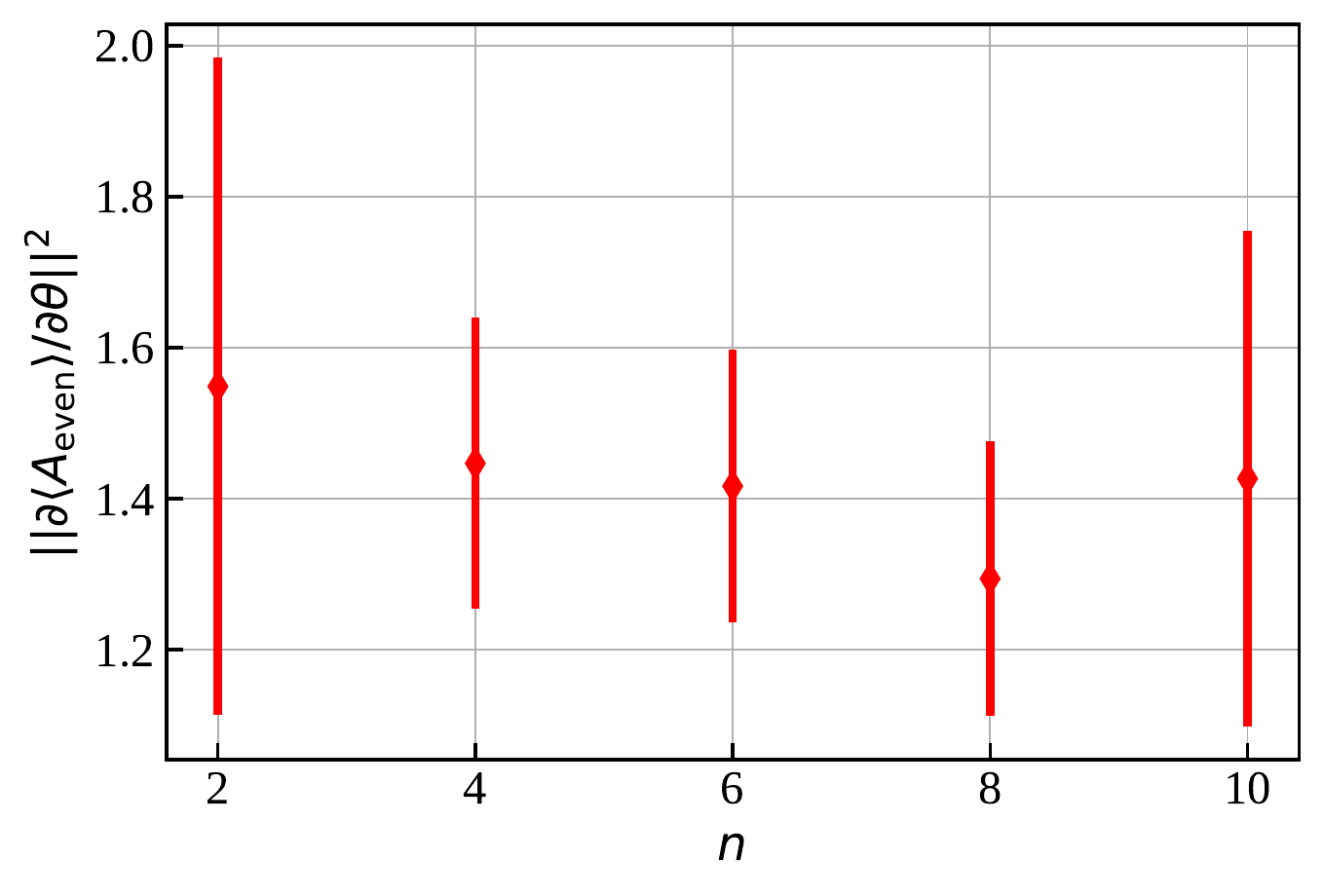}} \\
\subfloat[Norm of $\partial \Braket{A_\text{odd}} / \partial \boldsymbol{\theta}$]{\includegraphics[width=7cm]{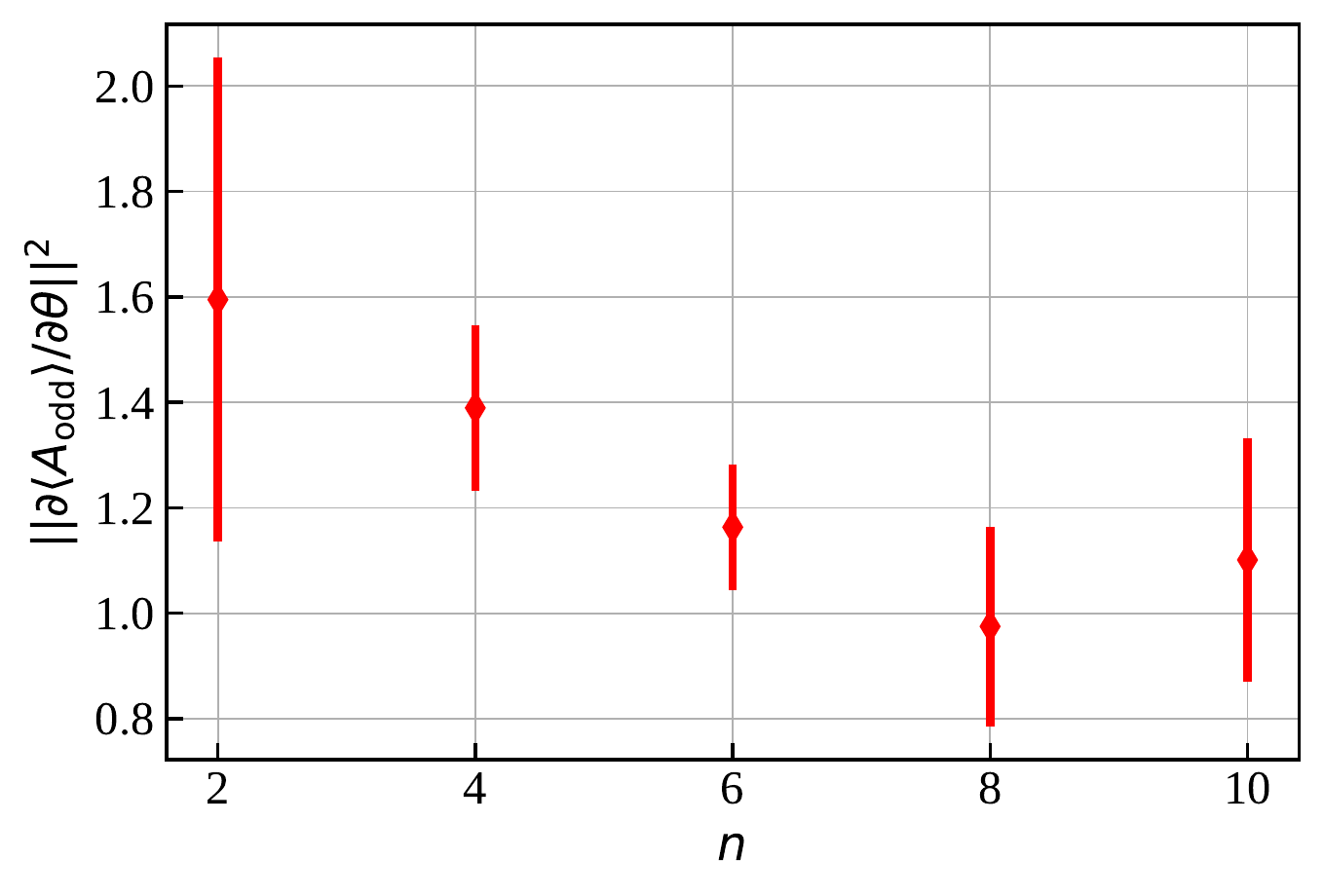}} 
\subfloat[Norm of $\partial \Braket{X \otimes I^{\otimes n}} / \partial \boldsymbol{\theta}$]{\includegraphics[width=7cm]{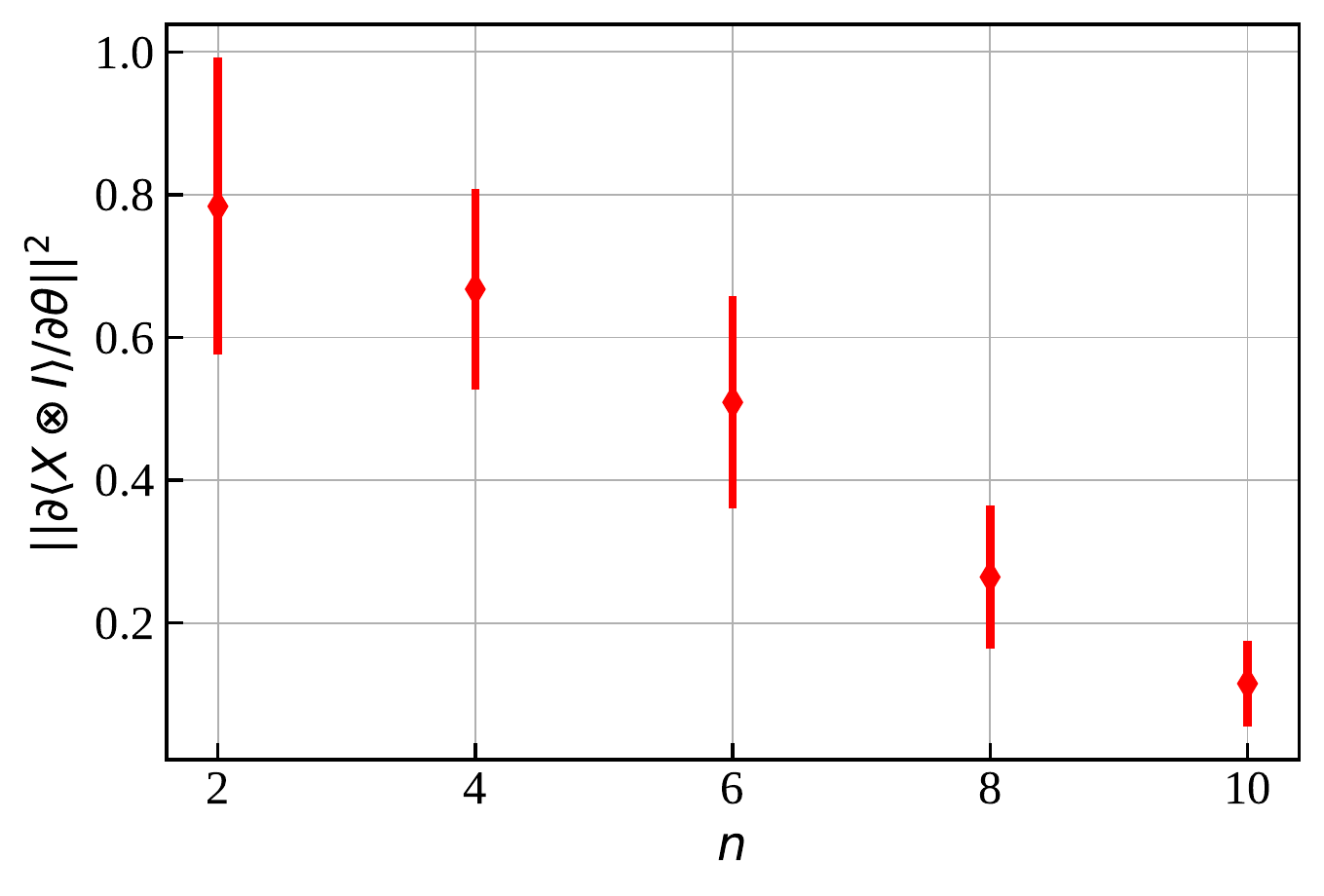}}
\caption{Norm of gradients of the cost function and those of each term composing the cost function, i.e., $\Braket{A_\text{even}} := \Braket{\psi(\boldsymbol{\theta}) | A_{\mathcal{T}_\text{even}} | \psi(\boldsymbol{\theta})}$, $\Braket{A_\text{odd}} := \Braket{\psi(\boldsymbol{\theta}) | A_{\mathcal{T}_\text{odd}} | \psi(\boldsymbol{\theta})}$, and $\Braket{X \otimes I} := \Braket{f, \psi(\boldsymbol{\theta}) | X \otimes I | f, \psi(\boldsymbol{\theta})}$. 
The points show the mean values of ten experiments from varying initial parameters and the error bars represent the standard deviation.\label{fig:grad_norm}}
\end{figure*}
%%%%%%%%%%%%%%%%%%%%%%%%
We examined the vanishing gradients of the cost function.
Figure \ref{fig:grad_norm} illustrates the L2-norm of the gradients of the cost function and those of each term composing the cost function, i.e., $\Braket{\psi(\boldsymbol{\theta}) | A_{\mathcal{T}_\text{even}} | \psi(\boldsymbol{\theta})}$, $\Braket{\psi(\boldsymbol{\theta}) | A_{\mathcal{T}_\text{odd}} | \psi(\boldsymbol{\theta})}$, and $\Braket{f, \psi(\boldsymbol{\theta}) | X \otimes I | f, \psi(\boldsymbol{\theta})}$.
These gradients were calculated using the \textit{statevector simulator}.
The number of layers of the ansatz was set to $5$.
The points show the mean values of ten experiments with different randomly set parameters and the error bars represent the standard deviations.
As shown in these figures, the gradient of the expectation of the operator $A_{\mathcal{T}_\text{even}}$, which is local, does not vanish, while the gradients of the other terms, which are global, vanish.
As a result, the gradient of the cost function as a whole vanishes.
The alleviation of barren plateaus will be addressed in future research.

\subsection{Dependency of the cost function error on the number of shots for the previous method} \label{sec:shoterror_ex}
\begin{figure*}[t]
\includegraphics[width=0.8\textwidth]{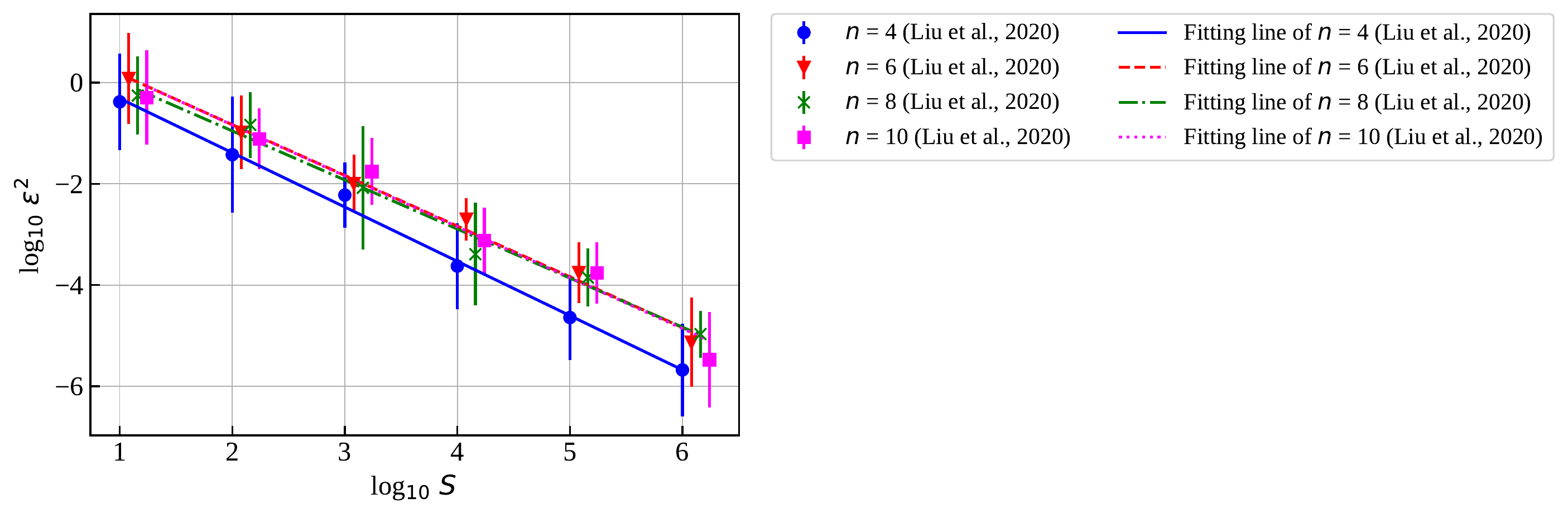}
\caption{Decadic logarithms of cost function error $\varepsilon$ vs. the number of shots $S$ for the previous method \cite{Liu2020variational} based on our implementation. The optimization was run ten times with fixed parameters for each number of shots. The mean values are plotted, with the error bars representing the standard deviations. The mean and standard deviation of the slopes of the fitted dashed line for $n=2, \ldots, 10$ (which includes cases that are not shown in this figure) are $-1.0$ and $0.05$, respectively. \label{fig:objective_qasm_previous}}
\end{figure*}

Figure~\ref{fig:objective_qasm_previous} shows the decadic logarithm of the squared error of the cost function with respect to the decadic logarithm of the number of shots for the previous method~\cite{Liu2020variational}.
In a similar fashion to our proposed method in Fig.~\ref{fig:objective_qasm_proposed}, the squared error decreases as the number of shots increases.
The mean and standard deviation of the slopes of the fitted lines are $-1.01$ and $0.04$, respectively.
Although a comparison of the proposed and previous methods from these figures is difficult because of the different definitions of the cost function, it can be deduced that the mean squared error in the cost function evaluation has a similar dependency on the number of shots in both methods.

% If you have acknowledgments, this puts in the proper section head.
\begin{acknowledgments}
This work is partly supported by UTokyo Quantum Initiative.
We thank Stuart Jenkinson, PhD, from Edanz Group (https://www.jp.edanz.com/ac) for editing a draft of this manuscript.
\end{acknowledgments}

% Create the reference section using BibTeX:
\bibliography{RefYSato}
\bibliographystyle{apsrev4-2}

\end{document}